%% file: cas-dc-template.tex
\documentclass[a4paper,fleqn]{cas-dc}

\usepackage[numbers]{natbib}
\usepackage{tabularray}
\usepackage{siunitx}

\def\tsc#1{\csdef{#1}{\textsc{\lowercase{#1}}\xspace}}
\tsc{WGM}
\tsc{QE}
\begin{document}
\let\WriteBookmarks\relax
\setcounter{topnumber}{5}
\setcounter{bottomnumber}{5}
\setcounter{totalnumber}{10}
\setcounter{dbltopnumber}{5}
\renewcommand{\topfraction}{0.95}
\renewcommand{\bottomfraction}{0.85}
\renewcommand{\textfraction}{0.05}
\renewcommand{\floatpagefraction}{0.8}
\renewcommand{\dbltopfraction}{0.95}
\renewcommand{\dblfloatpagefraction}{0.8}

\shorttitle{Gamma Backgrounds at HFIR}

\shortauthors{Heffron, B}

\title [mode = title]{Gamma Backgrounds for Experiments at the High Flux Isotope Reactor}  




%\author[1,2]{Blaine Heffron}[orcid=0000-0002-1564-2423]

\affiliation[1]{Brookhaven National Laboratory, Upton, NY, USA} \affiliation[2]{Department of Chemistry and Chemical Technology, Bronx Community College, Bronx, NY, USA} \affiliation[3]{Department of Physics, Boston University, Boston, MA, USA} \affiliation[4]{Department of Physics, Drexel University, Philadelphia, PA, USA} \affiliation[5]{George W. Woodruff School of Mechanical Engineering, Georgia Institute of Technology, Atlanta, GA, USA} \affiliation[6]{Department of Physics and Astronomy, University of Hawaii, Honolulu, HI, USA} \affiliation[7]{Department of Physics, Illinois Institute of Technology, Chicago, IL, USA} \affiliation[8]{Department of Physics, Le Moyne College, Syracuse, NY, USA} \affiliation[9]{Nuclear and Chemical Sciences Division, Lawrence Livermore National Laboratory, Livermore, CA, USA} \affiliation[10]{National Institute of Standards and Technology, Gaithersburg, MD, USA} \affiliation[11]{High Flux Isotope Reactor, Oak Ridge National Laboratory, Oak Ridge, TN, USA} \affiliation[12]{Physics Division, Oak Ridge National Laboratory, Oak Ridge, TN, USA} \affiliation[13]{Department of Physics and Astronomy, University of Tennessee, Knoxville, TN, USA} \affiliation[14]{Department of Physics and Astronomy, Johns Hopkins University, Baltimore, MD, USA} \affiliation[15]{Department of Physics, Susquehanna University, Selinsgrove, PA, USA} \affiliation[16]{United States Naval Academy, Annapolis, MD, USA} \affiliation[17]{Department of Physics, University of Wisconsin, Madison, WI, USA} \affiliation[18]{Wright Laboratory, Department of Physics, Yale University, New Haven, CT, USA}

\author[7]{M.~Andriamirado}[orcid=0000-0001-5459-4367]
\author[17]{A.~B.~Balantekin}[orcid=0000-0002-2999-0111]
\author[12]{C.~Baldenegro}
\author[8]{C.~D.~Bass}[orcid=0000-0001-9204-2280]
\author[7]{O.~Benevides~Rodrigues}[orcid=0000-0001-9181-6096]
\author[9]{E.~P.~Bernard}[orcid=0000-0002-2944-5359]
\author[9]{N.~S.~Bowden}[orcid=0000-0002-6115-0956]
\author[11]{C.~D.~Bryan}[orcid=0000-0001-6870-3608]
\author[16]{R.~Carr}[orcid=0000-0002-4181-5092]
\author[9]{T.~Classen}[orcid=0000-0001-6386-041X]
\author[11]{A.~J.~Conant}[orcid=0000-0001-7766-4321]
\author[4]{N.~Craft}
\author[11]{G.~Deichert}
\author[12]{A.~Delgado}[orcid=0000-0003-3453-7204]
\author[4]{M.~J.~Dolinski}[orcid=0000-0002-7716-2126]
\author[5]{A.~Erickson}[orcid=0000-0002-3276-4414]
\author[11]{M.~D.~Fuller}
\author[12,13]{A.~Galindo-Uribarri}[orcid=0000-0001-7450-404X]
\author[9]{S.~Ghosh}
\author[12,13]{C.~E.~Gilbert}
\author[12]{D.~C.~Glasgow}
\author[1]{S.~Gokhale}
\author[3]{C.~G.~Grant}[orcid=0000-0002-2328-1728]
\author[12,13]{B.~T.~Hackett}
\author[1,2]{S.~Hans}
\author[15]{A.~B.~Hansell}[orcid=0000-0003-2586-4019]
\author[10]{T.~E.~Haugen}[orcid=0000-0002-8222-116X]
\author[18]{K.~M.~Heeger}[orcid=0000-0002-4623-7543]
\author[12,13]{B.~A.~Heffron}[orcid=0000-0002-1564-2423]
\cormark[1]
\author[7]{A.~Irani}
\author[1]{D.~E.~Jaffe}[orcid=0000-0003-3122-4384]
\author[6]{J.~Koblanski}[orcid=0009-0008-3833-6921]
\author[4]{C.~E.~Lane}[orcid=0000-0003-4329-5796]
\author[7]{B.~R.~Littlejohn}[orcid=0000-0002-6912-9684]
\author[4]{A.~Lozano~Sanchez}
\author[12,13]{X.~Lu}
\author[7]{F.~Machado}[orcid=0000-0002-9436-0861]
\author[6]{J.~Maricic}[orcid=0000-0001-8431-8945]
\author[9]{M.~P.~Mendenhall}[orcid=0000-0002-5266-9940]
\author[6]{A.~M.~Meyer}[orcid=0009-0003-5304-5659]
\author[6]{R.~Milincic}
\author[12]{P.~E.~Mueller}[orcid=0000-0002-0334-6889]
\author[10]{H.~P.~Mumm}[orcid=0000-0003-4204-9817]
\author[4]{R.~Neilson}[orcid=0000-0002-2729-5131]
\author[12]{J.~R.~Newby}[orcid=0000-0003-3571-1067]
\author[14]{D.~Norcini}[orcid=0000-0003-0075-5326]
\author[6]{N.~Patel}
\author[9]{C.~Roca}[orcid=0000-0003-4994-5024]
\author[12,13]{E.~Romero-Romero}
\author[1]{R.~Rosero}
\author[12,13,14]{D.~Venegas-Vargas}[orcid=0000-0002-1548-8008]
\author[12]{B.~White}
\author[18]{J.~Wilhelmi}[orcid=0000-0002-5530-5130]
\author[1]{M.~Yeh}[orcid=0000-0003-2244-0499]
\author[1]{C.~Zhang}[orcid=0000-0003-2298-6272]
\author[9]{X.~Zhang}[orcid=0000-0003-2518-3651]

%\ead{baheffron@gmail.com}

%\ead[url]{https://blaineheffron.com}


%\affiliation[1]{organization={Oak Ridge National Lab},
%            addressline={1 Bethel Valley Road}, 
%            city={Oak Ridge},
%          citysep={}, % Uncomment if no comma needed between city and postcode
%            postcode={37830}, 
%            state={TN},
%            country={USA}}
%\affiliation[2]{organization={University of Tennessee},
%            city={Knoxville},
%          citysep={}, % Uncomment if no comma needed between city and postcode
%            postcode={37996}, 
%            state={TN},
%            country={USA}}

%\author[1,2]{Blaine Heffron}

% Footnote of the second author

% Email id of the second author
%\ead{}



%\affiliation[<aff no>]{organization={},
%%          citysep={}, % Uncomment if no comma needed between city and postcode

\cortext[1]{Corresponding author}



\begin{abstract}
    This article describes the deployment of a germanium detector at Oak Ridge National Lab's High Flux Isotope Reactor (HFIR) for the purpose of understanding the energy and spatial distribution of the gamma field in the experiment hall where the Precision Reactor Oscillation and Spectrum Experiment (PROSPECT) took data and future neutrino experiments could be located.
    The sources from both the reactor and the neutron beamlines are described in detail, along with their temporal variations due to reactor power and their spatial variations due to the
    geometry of the beamlines and building materials in the vicinity.
    Additionally, a shielding study was performed to assess the amount that backgrounds in tens of keV range can be mitigated. This work helps inform backgrounds for future experiments at reactors such as IBD-based neutrino measurements and CEvNS measurements.
\end{abstract}


% Research highlights are provided separately for submission and are not part of
% the NIM manuscript front matter.
%\item Characterization of gamma sources at ORNL's High Flux Isotope Reactor
%\item Spacial mapping of gamma backgrounds shows primary sources due to building geometry and neutron beamlines
%\item Shielding study for mitigating sources of gamma backgrounds at keV-scale energies 

% Keywords

\maketitle

% Main text
\include{paper_collab_review}

% Use the style of numbering in square brackets.
% If nothing is used, default style will be taken.
%\begin{enumerate}[a)]
%\item 
%\item 
%\item 
%\end{enumerate}  

%\begin{itemize}
%\item 
%\item 
%\item 
%\end{itemize}  

%\begin{description}
%\item[]
%\item[] 
%\item[] 
%\end{description}  

% Uncomment and use as the case may be
%\begin{theorem} 
%\end{theorem}

% Uncomment and use as the case may be
%\begin{lemma} 
%\end{lemma}


%\bibliographystyle{cas-model2-names}
\bibliographystyle{unsrt_mod}

\bibliography{cas-refs}

\end{document}

%% file: paper_collab_review.tex
\newcommand{\el}[2]{\ensuremath{^{#2}\textrm{#1}}}
\newcommand{\uFive}{\ensuremath{^{235}\textrm{U}}}
\newcommand{\cevns}{\ensuremath{\textrm{CE}\nu\textrm{NS}}}
\newcommand{\nue}{\ensuremath{\nu_{e}}}
\newcommand{\numu}{\ensuremath{\nu_{\mu}}}
\newcommand{\nuone}{\ensuremath{\nu_{1}}}
\newcommand{\nutwo}{\ensuremath{\nu_{2}}}
\newcommand{\nuthree}{\ensuremath{\nu_{3}}}
\newcommand{\nualpha}{\ensuremath{\nu_{\alpha}}}
\newcommand{\nubeta}{\ensuremath{\nu_{\beta}}}
\newcommand{\nuk}{\ensuremath{\nu_{k}}}
\newcommand{\nuj}{\ensuremath{\nu_{j}}}
\newcommand{\nutau}{\ensuremath{\nu_{\tau}}}
\newcommand{\nutaubar}{\ensuremath{\overline{\nu}_{\tau}}}
\newcommand{\nuebar}{\ensuremath{\overline{\nu}_{e}}}
\newcommand{\numubar}{\ensuremath{\overline{\nu}_{\mu}}}
\newcommand{\nubar}{\ensuremath{\overline{\nu}}}
\newcommand{\enubar}{\ensuremath{E_{\overline{\nu}}}}
\newcommand{\esmear}{\ensuremath{E_{smear}}}
\newcommand{\dcp}{\ensuremath{\delta_{CP}}}
\newcommand{\theonetwo}{\ensuremath{\theta_{12}}}
\newcommand{\theonethree}{\ensuremath{\theta_{13}}}
\newcommand{\thetwothree}{\ensuremath{\theta_{23}}}
\newcommand{\theonefour}{\ensuremath{\theta_{14}}}
\newcommand{\sinsqr}{\ensuremath{\sin^{2}\theta}}
\newcommand{\sinonetwo}{\ensuremath{\sin^{2}\theta_{12}}}
\newcommand{\sinonethree}{\ensuremath{\sin^{2}\theta_{13}}}
\newcommand{\sintwothree}{\ensuremath{\sin^{2}\theta_{23}}}
\newcommand{\sinonefour}{\ensuremath{\sin^{2}\theta_{14}}}
\newcommand{\mtwoone}{\ensuremath{\Delta m^2_{21}}}
\newcommand{\mthreeone}{\ensuremath{\Delta m^2_{31}}}
\newcommand{\mthreetwo}{\ensuremath{\Delta m^2_{32}}}
\newcommand{\mthreel}{\ensuremath{\Delta m^2_{3l}}}
\newcommand{\mfourone}{\ensuremath{\Delta m^2_{41}}}
\newcommand{\msqr}{\ensuremath{\Delta m^2}}
\newcommand{\msol}{\ensuremath{\Delta m^2_{\textrm{SOL}}}}
\newcommand{\matm}{\ensuremath{\Delta m^2_{\textrm{ATM}}}}
\newcommand{\mubest}{\ensuremath{\mu_{\textrm{best}}}}
\newcommand{\chisqr}{\ensuremath{\chi^2}}
\newcommand{\dchisqr}{\ensuremath{\Delta \chi^2}}
\newcommand{\dchisqrmin}{\ensuremath{\Delta \chi^2_{min}}}
\newcommand{\dchisqrcrit}{\ensuremath{\Delta \chi^2_c}}
\newcommand{\xasim}{\ensuremath{x^{Asimov}}}
\newcommand{\Tth}{\ensuremath{T_{\textrm{th}}}}
\newcommand{\kevee}{\ensuremath{keV_{ee}}}
\newcommand{\mevee}{\ensuremath{\textrm{MeV}_{\textrm{ee}}}}
\newcommand{\pspt}{PROSPECT}
\newcommand{\psptt}{PROSPECT-II}
\newcommand{\hfir}{\textsc{HFIR}}
\newcommand{\Geant}{\textsc{Geant}}
\providecommand{\g}{\ensuremath{\gamma}}
\newcommand{\gr}{$\gamma$-ray}

\section{Introduction}

The High Flux Isotope Reactor (HFIR) at Oak Ridge National Laboratory in Oak Ridge, TN USA is an 85 megawatt highly enriched uranium (HEU) reactor used by a wide array of scientists and industries as a neutron source. In the past two decades HFIR has also attracted interest from the particle physics community, primarily due to its status as a compact, intense emitter of antineutrinos\footnote{In this introduction, `neutrino' and `antineutrino' are used interchangeably where the context is clear.}. Neutrino experiments, such as \pspt~\cite{PROSPECT}, have detected HFIR's antineutrinos via the inverse beta decay (IBD) interaction, $\nuebar + p \rightarrow n + e^+$, to search for oscillations from new neutrino mass states and to measure the flux and energy spectrum of neutrinos produced by the daughters of \uFive~fission. Other neutrino detection efforts have considered HFIR as a site for performing fundamental science and applications-oriented neutrino measurements via the coherent elastic scattering channel, $\nu + X \rightarrow \nu + X$~\cite{coherent_first_det,coherent_proposed,HFIRWorkshop}. Still others have focused on HFIR as a potentially bright source of hard-to-detect hidden-sector particles not predicted by the Standard Model of Particle Physics~\cite{Sahoo:2024zee,Dent:2019ueq}.

A major challenge when operating an antineutrino detector at a reactor is the background associated with reactor operation, which comes primarily from neutrons that diffuse through the building into the area of the experiment. Thermalized neutrons can capture on building materials and experiment shielding materials, generating gamma radiation that can mimic neutrino detection signals if not properly shielded.

In this work, we describe results from deployment of a 1kg germanium (Ge) detector at the HFIR Experiment Hall, a wide corridor of the HFIR building located approximately 4 meters above the HFIR reactor, its neutron beamlines, and scattering instruments, that previously hosted the PROSPECT reactor antineutrino detector. These surveys provide a detailed characterization of the spatial and temporal variation of gamma radiation. The detector was deployed in two configurations: (a) a 'Russian doll' setup with concentric cylindrical layers of polyethylene and lead shielding to study achievable background reduction levels for future coherent elastic neutrino-nucleus scattering (CEvNS) measurements; and (b) a mobile lead collimator shield to survey gamma sources throughout the lab. Collimated studies characterized gamma sources from the reactor and beamline activities, including their dependence on reactor power and specific contributions from the HB4 beamline. Simulations of the collimated configuration enabled unfolding of gamma energy spectra to provide quantitative flux models—the primary product of this study. These models were validated against PROSPECT detector measurements to assess their utility for future experiments. Additionally, identified gamma lines may serve as in-situ calibration features for future experiments, enhancing measurement accuracy.

HPGe-based gamma characterizations are then compared to background measurements performed with the PROSPECT reactor antineutrino detector's 4-ton segmented liquid scintillator inner target. The reactor gamma background is characterized, and implications for future neutrino experiments are discussed.

\subsection{HFIR Building and Reactor Description}

The HFIR Experiment Hall is located approximately 4 meters above and 6 meters south of the reactor core. The pressurized reactor vessel is immersed in a pool of water which is surrounded by concrete. There are 4 neutron beam lines, two of which pass below the HFIR Experiment Hall - HB3 and HB4. HB3 is situated tangential to the reactor core and serves a Triple-Axis Spectrometer and the DEMAND Four-Circle Diffractometer~\cite{tax,demand}, while HB4 is also positioned tangentially and houses the cold neutron source that serves multiple instruments including the GP-SANS, Bio-SANS, CTAX, and IMAGINE diffractometers~\cite{ehlers2020}. These beamlines provide pathways for thermal neutrons to diffuse into the materials in the floor below the HFIR Experiment Hall.

The HFIR Experiment Hall contains a lead shield wall that is 0.1 m thick and surrounded by steel supports. The wall was constructed before the start of the \pspt~experiment to mitigate the largest source of backgrounds in the hall, which are gammas coming from the reactor pool wall. The central portion of the wall is 3.0 m wide and 2.1 m tall, completed by two shorter flanking portions on either end containing 0.05 m thick lead.

A thick concrete polygonal structure, hereafter referred to as the `concrete monolith' is positioned below the HFIR Experiment Hall's floor. It provides additional structural support and shielding around the reactor vessel. The Materials Irradiation Facility (MIF) valve box is attached to the wall just above the area where detector measurements were taken, containing residual radioactive contamination from past irradiation experiments.

\begin{figure*}[pos=htbp]
    \centering
    \includegraphics[width=1.0\linewidth]{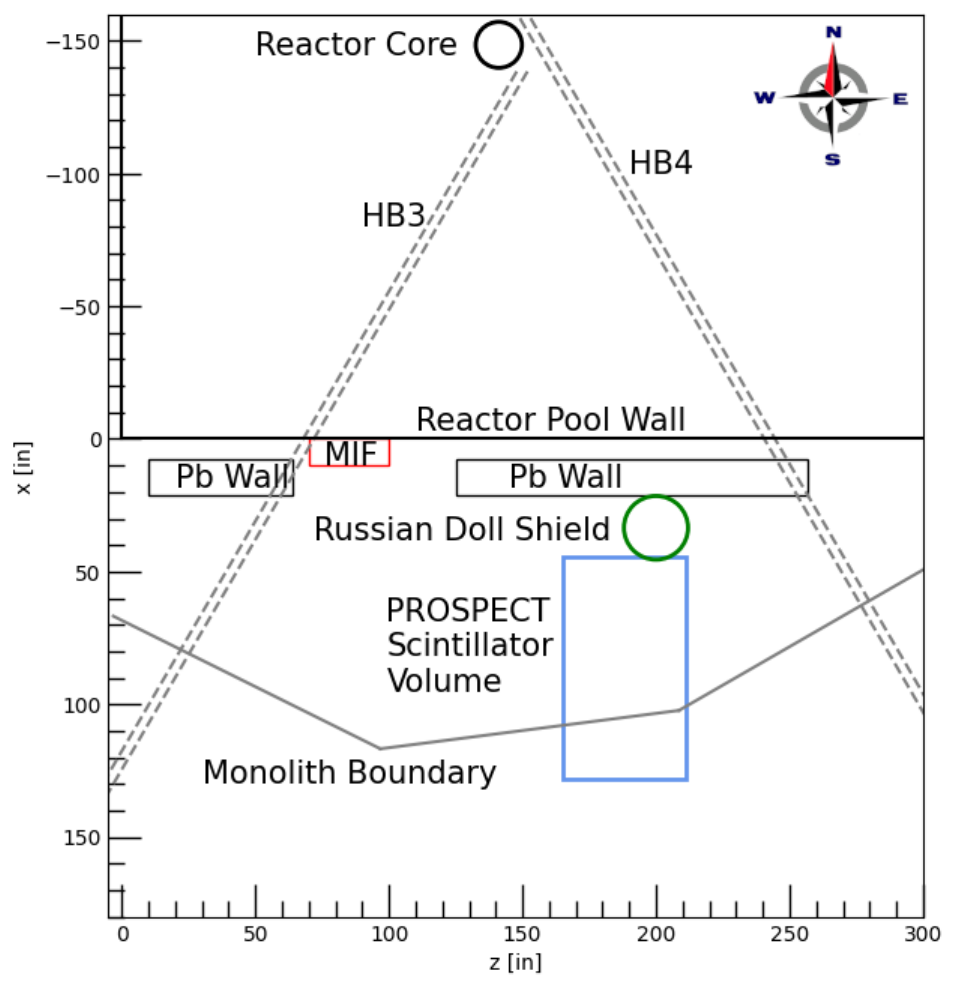}
    \caption{Top down diagram of the core and beamlines HB3 and HB4 relative to the HFIR Experiment Hall. Compass rose indicates directional orientation. The placement of the Russian Doll shield (refer to Section~\ref{sec:russian_doll} for more details) is located at the green circle at the lowest point of background radiation in front of the lead shield wall. The inner scintillator volume of where \pspt~took data is outlined in blue. The concrete monolith is outlined in grey. The location of the Materials Irradiation Facility (MIF) valve box is denoted in red. Note that the center of the core and the beamlines are 4 meters below the floor of the HFIR Experiment Hall.}
    \label{fig:hall_diagram}
\end{figure*}

Refer to Figure~\ref{fig:hall_diagram} for a diagram displaying the core and beamlines relative to features in the HFIR Experiment Hall, with axis labels indicating the coordinate system used throughout this work: x is the distance from the reactor wall, z is the distance from the western edge of the reactor wall, and y is the height above the floor.

\section{HPGe Survey Materials and Methods}

This section describes the HPGe detector and the two experimental configurations used for the gamma characterization measurements at the HFIR Experiment Hall.

The first is a lead collimator attached to a movable cart. 
The shield cart used is the ISOXSHLD produced by Mirion Technologies. It is movable with 4 iron caster wheels and uses a wheel lock rotational mechanism for rotating the shield and detector system up to 180 degrees. The lead collimator is cylindrical with a 111 mm outer radius, a 57 mm inner radius, and a 180 mm length. The lead end cap of length 43.5 mm contains a 30 degree opening cone at its center with an initial opening radius of 18 mm.

\begin{figure}[pos=htbp]
        \centering
        \includegraphics[width=0.8\columnwidth]{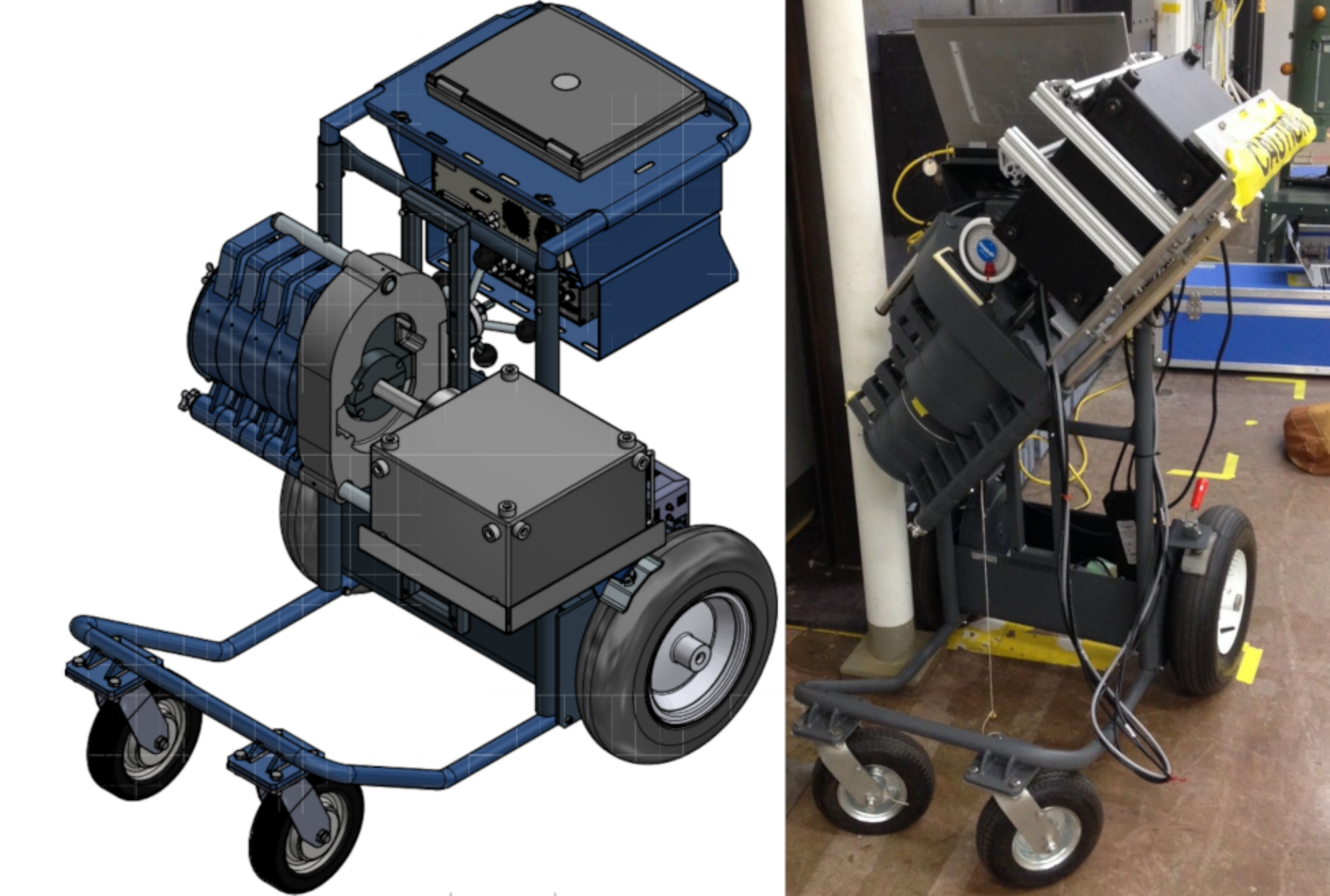}
        \caption{Left: isometric view of shield cart, right: picture of shield cart in position between the lead shield walls pointing towards the reactor core. In both views the collimator's lead endcap faces away from the viewer and is not visible.}
        \label{fig:shield_cart}
    \end{figure}

 Figure~\ref{fig:shield_cart} shows an isometric drawing of the shield cart with the detector placed in the horizontal position and a picture of the cart deployed at the reactor.  

This setup of the Germanium detector inside the lead collimator was used to scan throughout the room and point
to various sources of background. A catalogue of these measurements is reported in Section~\ref{sec:position_scan}.

The second configuration is a "Russian Doll" style arrangement of concentric cylindrical shielding surrounding the Ge detector. The shielding layers were chosen to mitigate thermal neutron and gamma backgrounds. Further description of the Russian Doll shielding will be given in Section~\ref{sec:rd-design}.

\subsection{Germanium Detector Specifications}
The detector is an Ortec™ GEM P-type coaxial High Purity germanium detector~\cite{ortec_gem_brochure}. The crystal has a 62~mm diameter by 69.7~mm length and a hole with an 8.5~mm diameter by 56.1~mm depth. The quoted dead layer thickness is 0.7~mm for the lithiated outer dead layer and a \SI{0.3}{\micro\metre} borated inner dead layer. The active volume calculated from the quoted dimensions is 1.02 ± 0.01~kg.

An aluminum case surrounds the detector and a copper cold finger is thermally coupled to the bottom of the crystal. The detector is cooled via an Ortec™ ICS mechanical cryo-cooler~\cite{ortec_ics_manual_2015}.

\subsection{Data Acquisition}

Signals are processed with a Canberra™ Lynx digital signal analyzer and analyzed using the Genie-2000 Spectroscopy Software from Canberra™~\cite{canberra_lynx_2008}. After a data-taking run, data files are output from GENIE-2000 in the form of a calibrated energy spectrum. Energy calibration is performed using standard radioactive sources (\el{Cs}{137} and \el{Na}{22}) with known gamma-ray energies, with calibrations verified before each measurement campaign.

A database was constructed to organize detector coordinates, data acquisition settings, detector bias voltage, shielding layouts, and energy calibrations used for each measurement. Data files were logically organized into runs which denote sequential measurements taken with the same detector configurations.

\section{Collimated Gamma Source Characterization}
\label{sec:reactor_gamma}

\subsection{Reactor Gamma Spectrum}

\begin{figure}[pos=htbp]
    \includegraphics[width=1.0\linewidth]{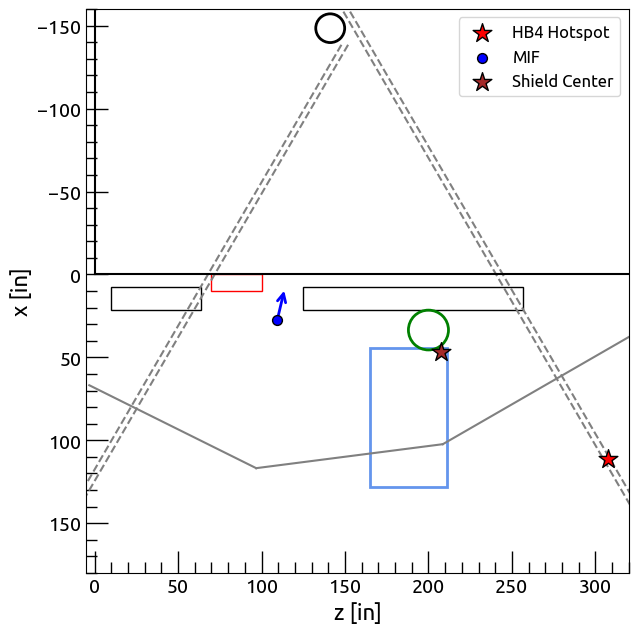}
    \caption{Location of the MIF reactor-facing position along with its orientation (blue arrow)~\ref{fig:reactor_spectrum}, Shield Center (brown star), and HB4 hotspot (orange star)~\ref{fig:HB4_shield}. Starred locations were measured with the collimator facing down.}
    \label{fig:key_measurement_locations}
\end{figure}

The HFIR Experiment Hall gamma backgrounds had previously been characterized with a germanium detector before the lead shield wall was erected for the PROSPECT experiment~\cite{2016prospectbackground}. This work builds on that previous effort with larger spatial coverage and a more detailed characterization of backgrounds due to beamline HB4. To measure gamma radiation coming from the reactor, the detector cart was placed next to the lead shield wall with the collimator facing the direction of the core, verified using facility mechanical drawings. The cart and detector angles were adjusted until the orientation with the maximum integral count rate was found.
Figure~\ref{fig:key_measurement_locations} shows the position of this measurement along with its orientation with the blue arrow, along with other key measurement locations detailed later in the paper (see Section~\ref{sec:beamline4} and Figure~\ref{fig:HB4_shield}).

\begin{figure}[pos=htbp]
    \includegraphics[width=1.0\linewidth]{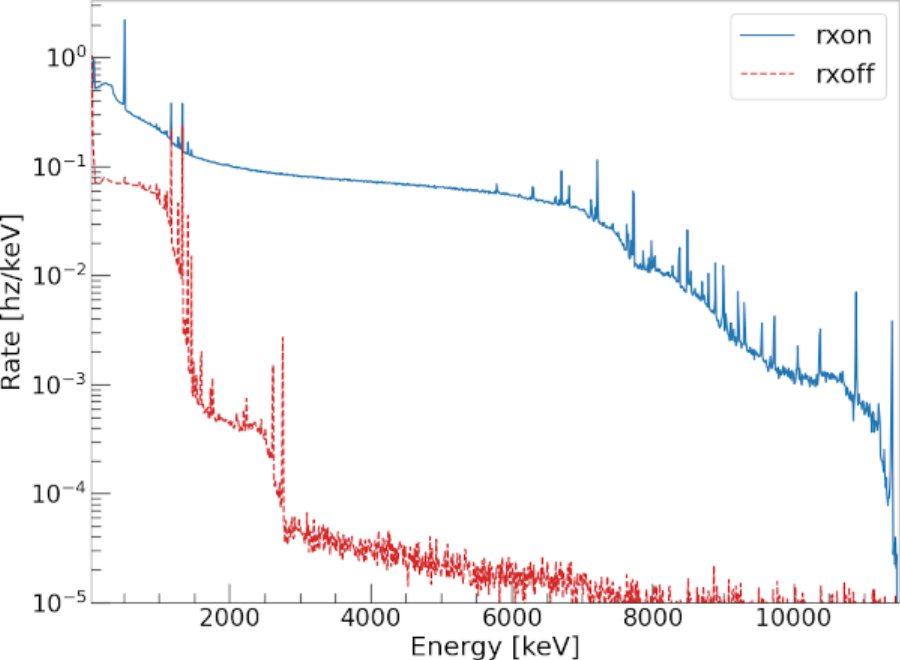}
    \caption{Measured energy spectrum with collimator facing the reactor positioned west of the lead shield wall, just under the MIF box (see figure~\ref{fig:key_measurement_locations}, blue arrow). Reactor-on in blue, reactor-off in red. See Table~\ref{table:source_catalogue} in Appendix~\ref{sec:spectral_lines} for a list of identified peaks.}
    \label{fig:reactor_spectrum}
\end{figure}

The detector was left in this position for 20 hours while the reactor was at 100\% operating power to obtain high statistics measurements. This measurement was repeated when the reactor was turned off. Figure~\ref{fig:reactor_spectrum} shows the measured spectra at this location.

Identified gamma spectral lines are listed in Table~\ref{table:source_catalogue} in Appendix~\ref{sec:spectral_lines}. Figures~\ref{fig:labelled_spectra_1}--\ref{fig:labelled_spectra_5} in Appendix~\ref{sec:spectral_lines} show the measured spectra with peaks labeled. Gamma lines above 5000 keV are due to neutron capture on isotopes present in structural steel throughout the building such as iron, nickel, and chromium~\cite{capgam}. Also visible are neutron capture lines from aluminum which forms the reactor vessel, and beryllium from the neutron reflector surrounding the core.

In the 1-2 MeV range there are gammas from beta decaying radioactive isotopes such as $^{60}$Co, $^{152}$Eu, $^{154}$Eu, $^{41}$Ar and $^{40}$K~\cite{nudat}.
The Europium is present in the reactor pool due to the control plates  composed of $\textrm{Eu}_2\textrm{O}_3$. 
Cobalt is present from residual contamination from past experiments at the Materials Irradiation Facility (MIF) 
which has a valve box attached to the wall just above where the detector was placed. 
$^{41}\textrm{Ar}$ circulates through the building while the reactor is operational due to neutron capture on $^{40}\textrm{Ar}$ present in the air around the pool.

Sources related to residual activity of radioactive isotopes from reactor activities are also present when the reactor is off. Notably, the 2754 keV line from \el{Na}{24} beta decay and 2614 keV line from \el{Tl}{208} beta decay are visible in the reactor-off spectrum but are not strong enough to be visible above background while the reactor is running.  $^{24}\textrm{Na}$ is present in building materials activated from neutron capture on $^{23}\textrm{Na}$ while the reactor is on. This isotope has a 15 hour half life so its activity decays over a few days after the reactor shuts down.
Building materials contain thorium and its decay products including \el{Tl}{208}. 
See table~\ref{table:source_catalogue} for a list of all gamma lines and their sources. 

\subsection{Gamma Rates Correlated with Reactor Power}
\label{sec:rate_power_correlation}

\begin{figure}[pos=htbp]
    \centering
    \includegraphics[width=1.0\linewidth]{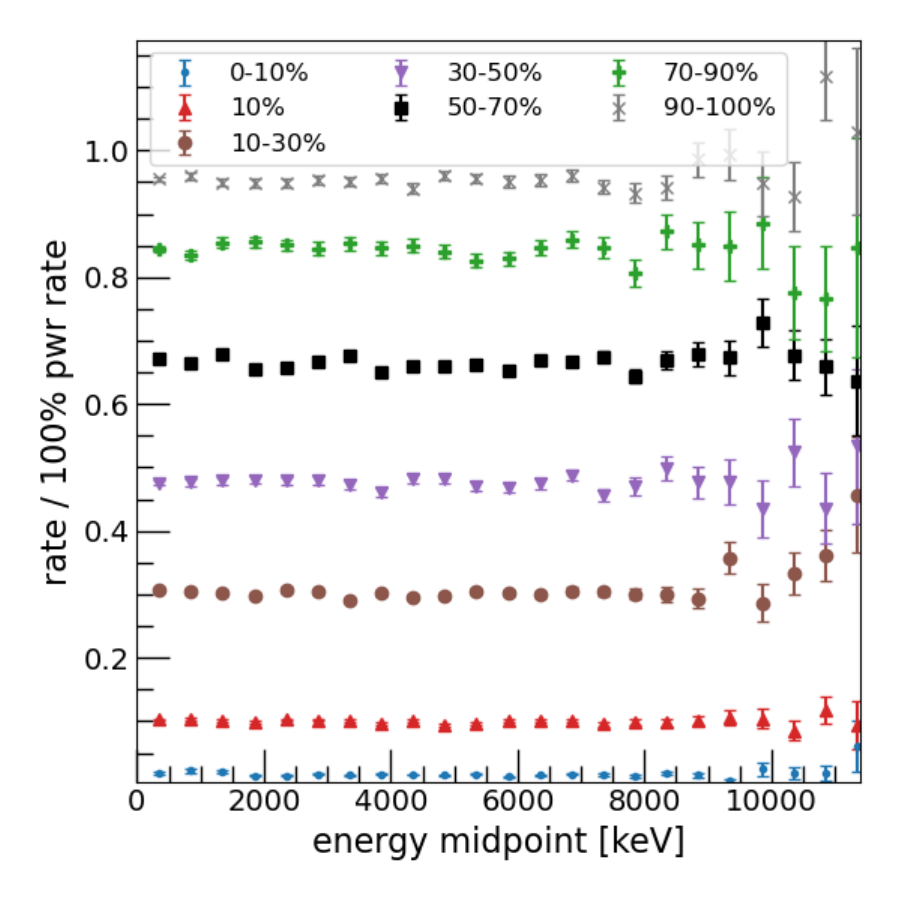}
    \caption{Ratios of rates in 500 keV increments to the spectrum at 100\% power for various periods during the ramp up for cycle 491. Note that only power ranges are known based on the operator announcing the beginning of a ramp up to a set power level with the exception of the 10\% power level which the reactor was kept at for the duration of the 10\% sample.}
    \label{fig:rate_ratios}
\end{figure}

The measured rate of gamma rays in the HFIR Experiment Hall correlate strongly with reactor power. During reactor startup for cycle 491, spectra were collected at each power level as announced by operators. The detector was positioned in the same location and orientation as was used for the spectrum taken in Figure~\ref{fig:reactor_spectrum}. A reactor off spectrum taken over 5 days before startup was subtracted from measured spectra. Figure~\ref{fig:rate_ratios} shows that various portions of the gamma energy spectrum correlate well with reactor power, with rates at 10\% power showing approximately 10\% of the rates observed at 100\% power, confirming the expected linear relationship.
No trends in the energy spectrum emerge at different power levels, indicating that the entire energy range is fully correlated with reactor power.

\subsection{Gamma Rates Time Series}
\label{sec:rate_time_series}
\begin{figure}[pos=htbp]
    \centering
    \includegraphics[width=1.0\linewidth]{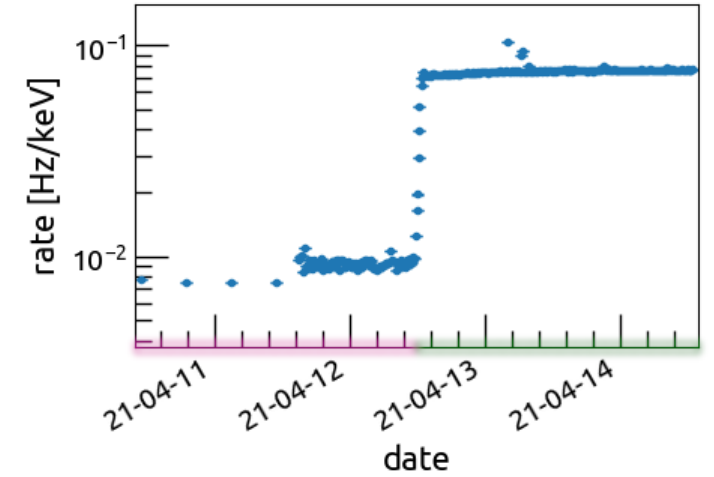}
    \caption{Rates between 30 keV and 11465 keV over time in the collimated Ge detector while in the MIF reactor-facing position (see Figure~\ref{fig:key_measurement_locations}, blue arrow). Reactor-off (on) time period highlighted in red (green).}
    \label{fig:rate_time_series}
\end{figure}

During the 2 day lead up to Cycle 491 and the first 2 days of the cycle, the detector was placed in the MIF reactor-facing position.
Before this particular cycle some reactor tests were performed the day before the startup. This involved bringing the reactor up and down between 0 and 10\% power which is illustrated in the rates observed in Figure~\ref{fig:rate_time_series}.

One can see that on April 13, 2021 between 05:00 and 10:00 the rates spiked at certain times. This is due to instrumentation in beamline 3 below the floor changing settings causing larger amounts of neutrons to scatter up into the experiment hall. A neutron diffractometer~\cite{demand} and spectrometer~\cite{tax} are instruments capable of these changes in neutron flux observable in the experiment hall~\cite{Hackett2017DANGAT,Heffron2017,Heffron2023}. The $\sim$30\% variation seen in this plot is typical of such variations seen during a reactor cycle.

The observed rate variations, particularly the spikes on April 13, highlight the influence of beamline activities on ambient neutron flux in the experiment hall.
Future experiments requiring precise background characterization, such as \cevns~measurements, should implement real-time monitoring of ambient backgrounds to account for transient changes from beamline operations and reduce systematic uncertainties.

\section{Spatial Survey of Gamma Sources}
\label{sec:position_scan}

Based on our understanding of the building features such as the shape of the concrete monolith, the location of the beamlines and associated beam shutters, and the volume of the reactor pool, we hypothesized there would be two primary regions of elevated gamma flux relevant to the experiment hall detector area. These correspond to the region above the HB4 beamline and the radiation coming through the north wall near the reactor pool. A thorough scan was performed to test this hypothesis and characterize the spatial variation of the gamma field in order to better understand radiation sources.

\subsection{Downward-Facing Survey Results}

\begin{figure*}[pos=htbp]
    \centering
    \includegraphics[width=0.8\linewidth]{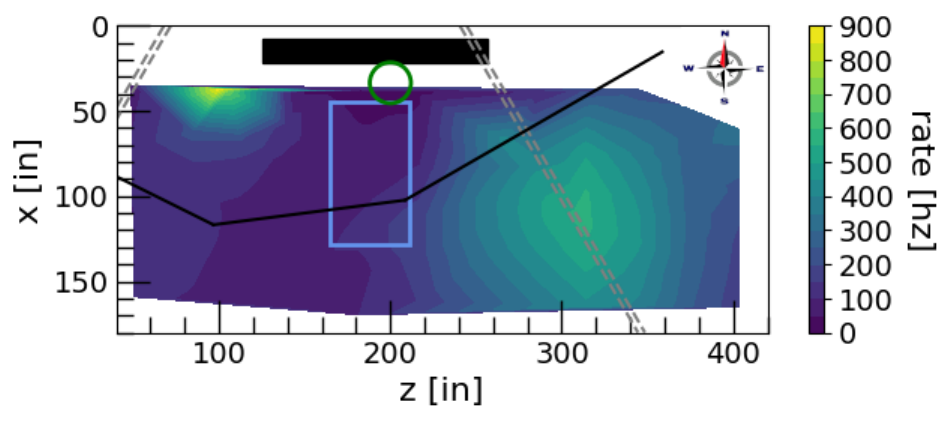}
    \caption{Rates between 50 keV and 11400 keV measured with the Ge detector inside the lead collimator facing down towards the floor. Blue box indicates the region of the \pspt~active scintillator volume, the green circle represents the location of the Russian doll shield, the grey dashed lines represent the path of beamline 4 through the floor below, and the black rectangle represents the lead shield wall.
        The black line denotes the boundary of the concrete monolith positioned below the floor.
        The x = 0 position is the reactor wall, coordinates are in inches. The region of high rate around z=300 in, x=110 in corresponds to shielding surrounding the cold source of the HB4 beamline. The region of high rate at x=40 in, z=100 in corresponds to radiation coming from the reactor without the lead shield wall.}
        
    \label{fig:down_scan}
\end{figure*}

Approximately 100 positions were scanned with 4 minutes of live time using a low gain mode allowing an energy range up to 11.4 MeV with the collimator facing directly down towards the floor at a height of 41.5 inches above the floor. This orientation was chosen to get a uniform exposure in the horizontal dimensions. A triangulated contour plot showing total rates for energies from 30 keV to 11400 keV is shown in Figure~\ref{fig:down_scan}. The highest rates are found west of the lead shield wall where radiation from the reactor is least attenuated. Relative to the unshielded western side of the map in Figure~\ref{fig:down_scan}, the lead shield wall reduces total gamma backgrounds in the region of \pspt~by approximately a factor of 5-10. The mitigation is even more pronounced at higher energies, with a 60-fold reduction from 4 to 7 MeV and an 80-fold reduction at energies greater than 7 MeV. A hot spot is visible east of the \pspt~location which corresponds to large amounts of steel surrounding the cold source within the HB4 beamline on the floor below. This hotspot is the dominant source of gamma backgrounds for the \pspt~experiment based on correlation with elevated backgrounds in the corresponding detector segments~\cite{PROSPECT}.

\begin{figure}[pos=htbp]
    \centering
    \includegraphics[width=1.0\linewidth]{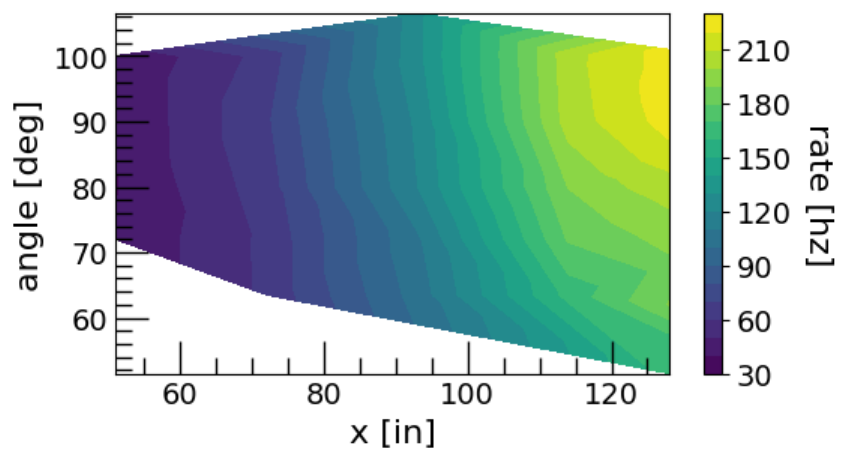}
    \caption{Measured gamma rates between 50 and 11400 keV. Scan is along the east face of the \pspt~detector active volume. Low x values correspond to the north side of the detector. High rates correspond to the detector being pointed at the HB4 hot spot. }
    \label{fig:east_scan}
\end{figure}

Another scan was done along the east face of the position of the \pspt~inner volume with the detector pointed east towards the HB4 hotspot. This was done to quantitatively determine the gamma rate contribution to the \pspt~detector from the hotspot. The rates from the scan are depicted in Figure~\ref{fig:east_scan}. An angle of 0 degrees corresponds to the detector pointing directly downward. Simulation of the gamma backgrounds in the \pspt~detector are detailed in Section~\ref{sec:prospect_gammas}.

A similar scan was attempted along the $z$ direction pointing towards the reactor to see if we could locate a spike in rate correlated with the reactor location. This yielded statistically insignificant results due to low per-run statistics.

\subsection{Beamline 4 Gamma Characterization}
\label{sec:beamline4}

As shown in Figure~\ref{fig:hall_diagram}, beamline HB4 runs East of the \pspt~location at a 30 degree angle counter clockwise from south through the floor below. As can be seen in the position scan in Figure~\ref{fig:down_scan}, there is a region of larger gamma rates roughly 100 inches East of the southern edge of the \pspt~detector liquid scintillator volume. 
The center of this hot spot corresponds to a protective steel block surrounding the beamline at the cold source of beamline HB4. This cold source is a cryogenic moderator system that uses supercritical hydrogen to slow thermal neutrons into lower-energy cold neutrons for scattering experiments; the steel block serves primarily as radiation shielding to attenuate scattered neutrons and gamma rays, while also providing mechanical containment for potential cryogenic hazards.
This source of radiation caused larger amounts of gamma backgrounds in the bottom two layers of segments on the south side of the \pspt~detector, as established in previous work~\cite{PROSPECT2018, PROSPECT}.

\begin{figure}[pos=htbp]
    \centering
    \includegraphics[width=1.0\linewidth]{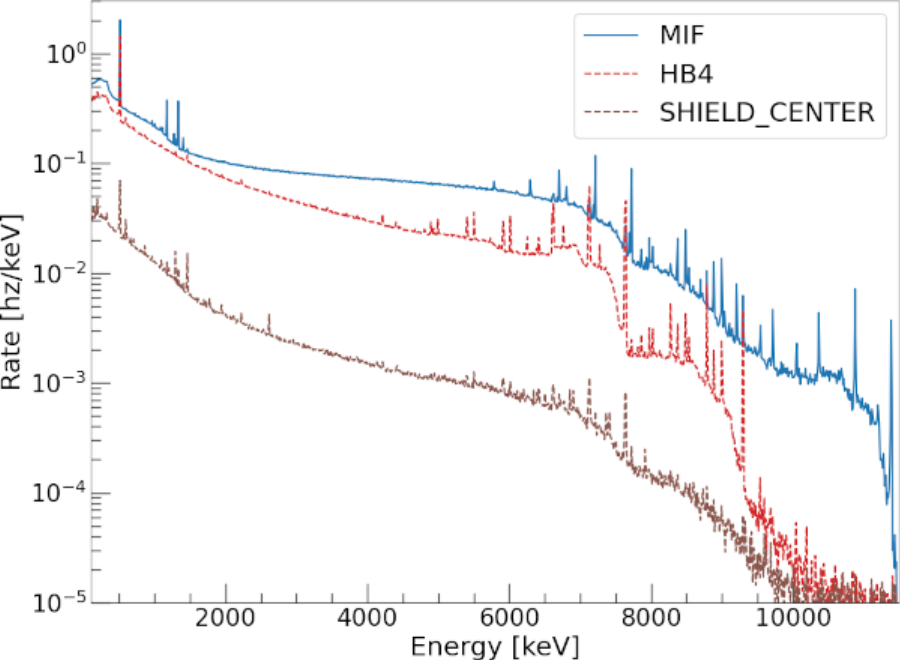}
    \caption{Comparison of reactor-on spectra taken at three different locations. Blue is taken at the MIF reactor-facing position with collimator pointing at the reactor (no lead shield wall attenuating this location), red is taken over the HB4 hot spot, and brown is at the Shield Center location situated close to the center of the lead shield wall. See Figure~\ref{fig:key_measurement_locations} for each measurement position.}
    \label{fig:HB4_shield}
\end{figure}

An overnight run was made with the detector placed at the center of the hot spot with the detector pointed down to obtain a high statistics energy spectrum. This spectrum is shown in Figure~\ref{fig:HB4_shield} along with spectra taken facing the reactor both through and beside the lead shield wall for comparison (see Figure~\ref{fig:key_measurement_locations} for positions). The major difference between the HB4 radiation and the reactor gammas are the lack of reactor-specific sources such as neutron capture on beryllium, aluminum, and $^{59}$Ni and the presence of neutron capture on $^{48}$Ti. The presence of titanium, inferred from the observed spectral lines, suggests different steel composition in the beamlines compared to structural steel surrounding the reactor.

For a future IBD-based neutrino experiment at HFIR, mitigating HB4 gammas would reduce accidentals in active detector areas above the back edge of the concrete monolith. Other above- or near-monolith locations adjacent to the HFIR Experiment Hall may also achieve similarly acceptable gamma background levels by reproducing the \pspt~lead shield wall structure.

\section{Russian Doll Shield Measurements}
\label{sec:russian_doll}

The primary purpose of the Russian doll (RD) configuration was to examine backgrounds at low energy relevant for \cevns~detection. 

\subsection{Russian Doll Shielding Design}
\label{sec:rd-design}

The Russian doll shield was constructed by AMETEK-AMT; it consists of a series of concentric cylindrical shields. From outer to inner there is a half inch thick steel casing, 3.5" of lead, 0.02" tin, 1/16" copper, 2 7/8" borated polyethylene and 0.9" lithiated polyethylene at the innermost layer. The tin and copper layers attenuate x-rays from the lead. Borated polyethylene has a high thermal neutron capture cross-section due to the $^{10}$B(n,$\alpha$)$^{7}$Li reaction. The inner lithiated polyethylene layer was chosen to absorb neutrons via the $^{6}$Li(n,$\alpha$)$^{3}$H reaction which does not produce gammas reaching the germanium detector. A comparison between shielding materials used for the RD and PROSPECT is shown in table~\ref{tab:rd_compare}.

The shielding sits atop a steel stand (not pictured) with the center of the HPGe positioned 41.5 inches above the ground. A layer of water bricks was added around the shield apparatus for some measurements. Each water brick contains 3.5 gallons of water and has dimensions 9" × 18" × 6". The water brick layer extends 1 brick thick around the perimeter. We took measurements both with and without water bricks to assess their impact on thermal neutron backgrounds.

\renewcommand\cellalign{ll}
\begin{table*}[pos=!tbp]
\centering
\caption{Approximate comparison of shielding layers in PROSPECT~\cite{PROSPECT2018} and the Russian Doll configuration (innermost to outermost). The PROSPECT entries are intended only as representative design-scale values for qualitative comparison with the RD assembly.}
\label{tab:rd_compare}
\begin{tabular}{|c|l|c|l|c|}
\hline
Layer & \multicolumn{1}{c|}{PROSPECT Material} & \multicolumn{1}{c|}{Thickness (m)} & \multicolumn{1}{c|}{Russian Doll Material} & \multicolumn{1}{c|}{Thickness (m)} \\
\hline
1 & Acrylic tank wall & 0.063 & Lithiated polyethylene & 0.023 \\
\hline
2 & Water & 0.025 & Borated polyethylene & 0.073 \\
\hline
3 & Borated polyethylene & 0.025--0.075 & Copper & 0.0016 \\
\hline
4 & Outer aluminum tank wall & 0.025 & Tin & 0.00051 \\
\hline
5 & Lead & 0.025 & Lead & 0.089 \\
\hline
6 & Structural polyethylene timbers & 0.10 & Steel casing & 0.013 \\
\hline
7 & Borated polyethylene & 0.025 & Water bricks (optional) & 0.23 \\
\hline
8 & Outer aluminum skin & 0.000762 & -- & -- \\
\hline
\end{tabular}
\end{table*}

%\begin{landscape}
%\\begin{SidewaysFigure}
%\begin{figure}[pos=htbp]
%  \centering
%    \includegraphics[width=1.0\columnwidth]{figures/pdf/HFIRBG_DB.png}
%    \caption{Database schema used for the experiment. Lines denote foreign key relationships.}
%    \label{fig:db}
%\end{figure}
%\end{SidewaysFigure}
%\end{landscape}

\begin{figure*}[pos=htbp]
    \centering
    \includegraphics[width=1.0\linewidth ]{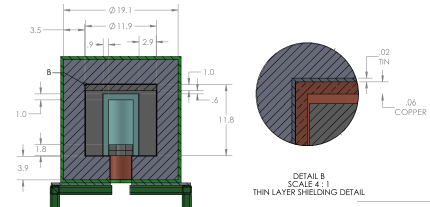}
    \caption{Cutout drawing of Russian Doll shielding.
        From outer to inner shielding layers are steel (green), lead (light grey), tin (see detail B), copper (see detail B), borated polyethylene (dark grey) and lithiated polyethylene (teal). The brown cylinder below the detector is the opening in the lead for the copper cold finger and cryocooler housing to insert through. Measurements in inches.}
    \label{fig:russian_doll}
\end{figure*}

\begin{figure}[pos=htbp]
    \centering
    \includegraphics[width=1.0\linewidth]{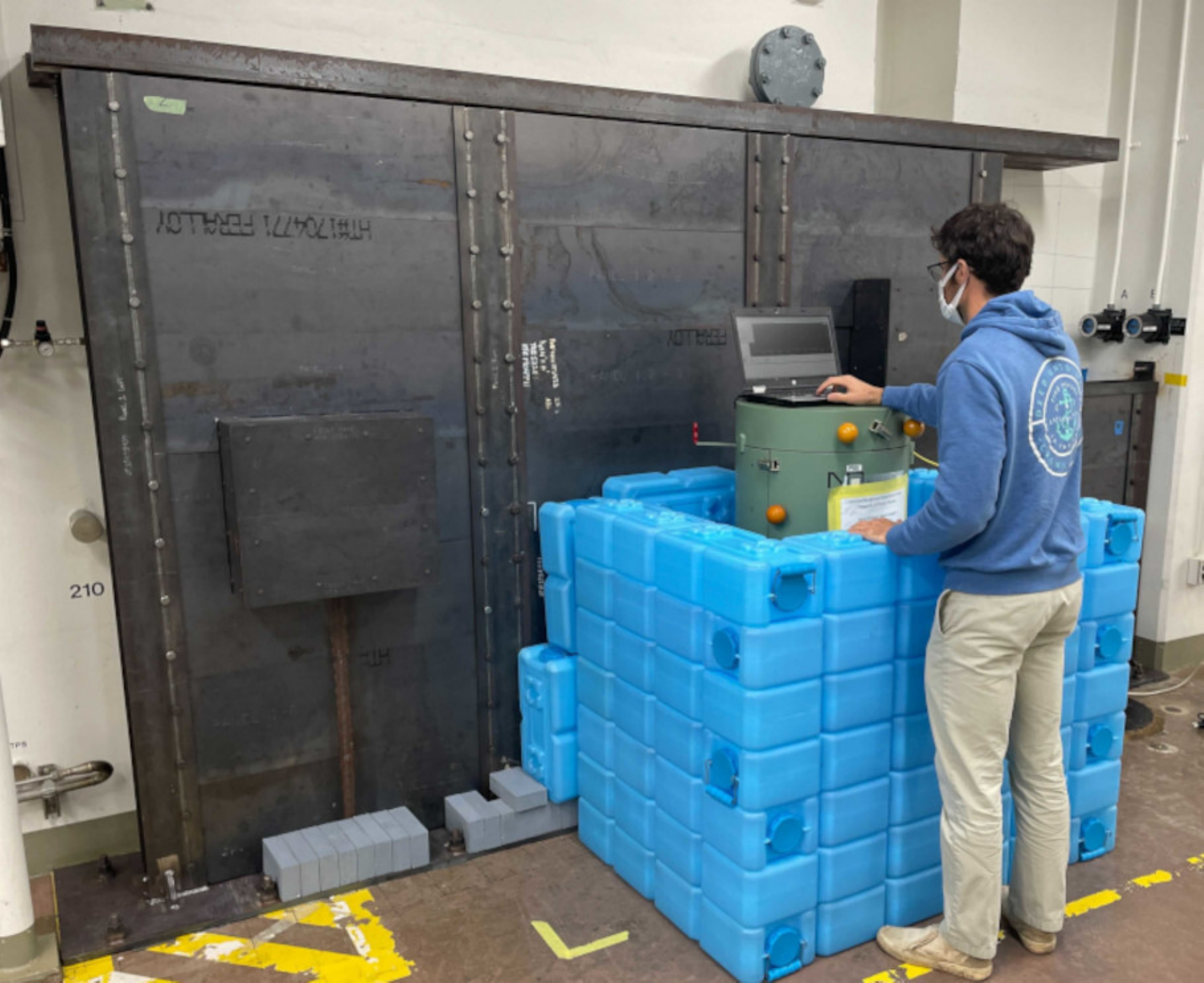}
    \caption{Russian Doll shielding water bricks. Background shows the lead shield wall.}
    \label{fig:rd_water}
\end{figure}

\begin{figure*}[pos=htbp]
    \centering
    \includegraphics[width=1.0\linewidth]{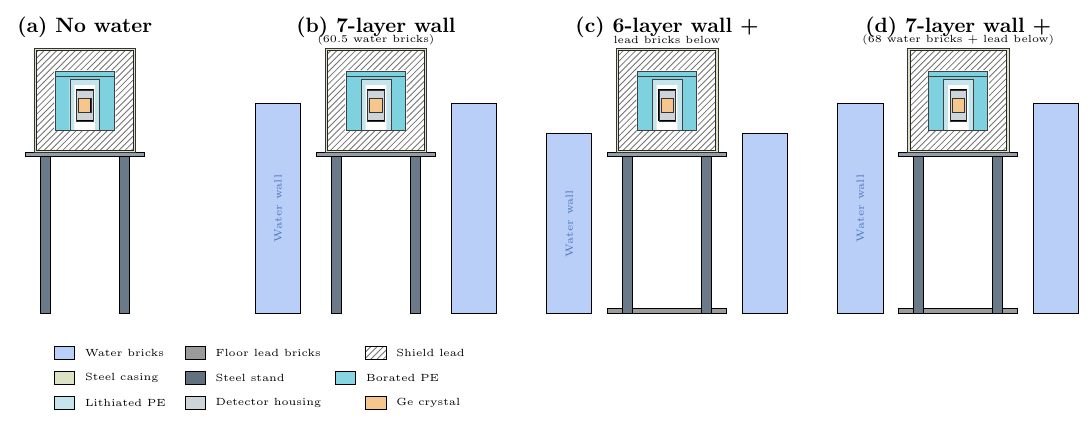}
    \caption{Scaled front-elevation schematics of the four shielding configurations used with the Russian Doll shield. Water bricks formed an open-top square wall around the full RD assembly, including the stand, with a small gap between the wall and the stand legs. The lead-under-cart configurations used a flat tiled layer of 0.025$\times$0.1$\times$0.2~m lead bricks directly beneath the cart. (a) Baseline RD configuration with no additional shielding. (b) 7-layer water-brick wall (60.5 water bricks total). (c) 6-layer water-brick wall with lead plate under the cart. (d) 7-layer water-brick wall with lead plate under the cart (68 water bricks total). Each water brick has dimensions 9$\times$18$\times$6~inches.}
    \label{fig:water_brick_configs}
\end{figure*}

\subsection{Russian Doll Measurement Campaigns}

After taking various measurements with the movable collimator configuration, the germanium detector was placed in the RD shielding in July, 2021 where it took data until February, 2022 during which 4 reactor on/off cycles took place. 
Two gain configurations were used during the runs, one the same gain which was used to measure up to 11.5 MeV as shown with the collimator scans. The other gain was a high gain mode used to examine the energy space below 90 keV. This was
done to examine backgrounds at energies relevant for \cevns~detection.
Beyond the RD shielding described in Section~\ref{sec:russian_doll}, some runs featured an additional layer of lead 0.025~m thick placed below the shield. This was accomplished with a 3$\times$4 arrangement of 0.025$\times$0.1$\times$0.2~m bricks.
We also took data-taking runs with a variety of water brick shielding configurations, which will be described later in Section~\ref{sec:rd_shield_compare}.

See table~\ref{table:rate_comparison} for a comparison between detected rates in the RD configuration (no additional lead or water bricks) and the collimator configuration at different locations for energies from 100~keV to 11.5~MeV. By comparing this dataset to that taken with the bare collimator at the nearby ‘shield center’ location, we can see that the RD shielding reduces detected gamma rates by 55\%. 

\renewcommand\cellalign{ll}
\begin{table*}[pos=!tbp]
\caption{Overall rates in the energy range 100 keV to 11.5 MeV for reactor-on and reactor-off periods. Location labels correspond to the measurement positions shown in the referenced figures: MIF is the reactor-facing position under the MIF valve box, Shield Center is near the center of the lead shield wall, Russian Doll is the concentric shielding configuration, and HB4 Hotspot is above the beamline 4 cold source.}
\centering
\begin{tabular}{llll}
\toprule
Location & Reactor-on (mHz/keV/kg) & Reactor-off (mHz/keV/kg) & Reactor-only (mHz/keV/kg) \\
\midrule
MIF~\ref{fig:reactor_spectrum} & 77.2 $\pm$ 3.9 & 6.74 $\pm$ 0.34 & 70.5 $\pm$ 3.9 \\
Shield Center~\ref{fig:HB4_shield} & 3.28 $\pm$ 0.16 & 1.11 $\pm$ 0.06 & 2.17 $\pm$ 0.17 \\
Russian Doll~\ref{fig:RD_on_off_compare} & 1.48 $\pm$ 0.07 & 0.275 $\pm$ 0.014 & 1.21 $\pm$ 0.07 \\
HB4 Hotspot~\ref{fig:HB4_shield} & 46.9 $\pm$ 2.3 & 0.603 $\pm$ 0.030 & 46.3 $\pm$ 2.3 \\
\bottomrule
\end{tabular}
\label{table:rate_comparison}
\end{table*}

\subsection{Primary Gamma Sources within the Russian Doll}
\label{sec:rd_sources}

Due to the presence of the nearby lead shield wall and to the large amount of lead in the RD shield, most gamma sources present in the HFIR Experiment Hall are attenuated to levels where no peaks are visible in the spectrum. Additionally, the borated and lithiated polyethylene shielding attenuate thermal neutrons from the reactor before they can capture on the germanium. At these levels of attenuation with the baseline RD configuration, gamma backgrounds are about one order of magnitude away from achieving a detector environment with comparable reactor-on and reactor-off background rates.

The primary backgrounds remaining are gammas from neutrons capturing on the shielding materials, natural radioactivity in the shielding materials, the radon decay chain from radon in the air, and gammas down-scattering off the shield into the detector.

The main lines visible are listed in table~\ref{tab:rd_lines}. $^{113}$Cd is very prevalent in the spectrum. This was unexpected since no cadmium-bearing material was listed in the shielding specifications provided by AMETEK. To estimate the order of magnitude of cadmium-bearing material required to reproduce this line strength, a simplified Geant4 study was performed in which natural cadmium was added phenomenologically to the tin layer. This should be interpreted only as a modeling ansatz for the observed line strength, not as an identification of the specific component responsible.
To do this an isotropic thermal neutron flux was simulated in Geant4~\cite{geant4-1,geant4-2,geant4-3}. The cadmium fraction in that phenomenological tin-layer model was then varied so that the
ratio of the area of the neutron capture peak on \el{Pb}{207} (7368 keV) to the area from the primary \el{Cd}{113} line (558 keV) matched the data.
Within that simplified model, the best match corresponds to $\approx$15\% of the atoms in the tin mixture being natural cadmium, indicating roughly 2\% is \el{Cd}{113}. Given the simplified neutron field and the unknown material origin of the cadmium line, this should be regarded only as an order-of-magnitude estimate rather than a materials assay.

\begin{figure}[pos=htbp]
    \centering
    \includegraphics[width=1.0\linewidth]{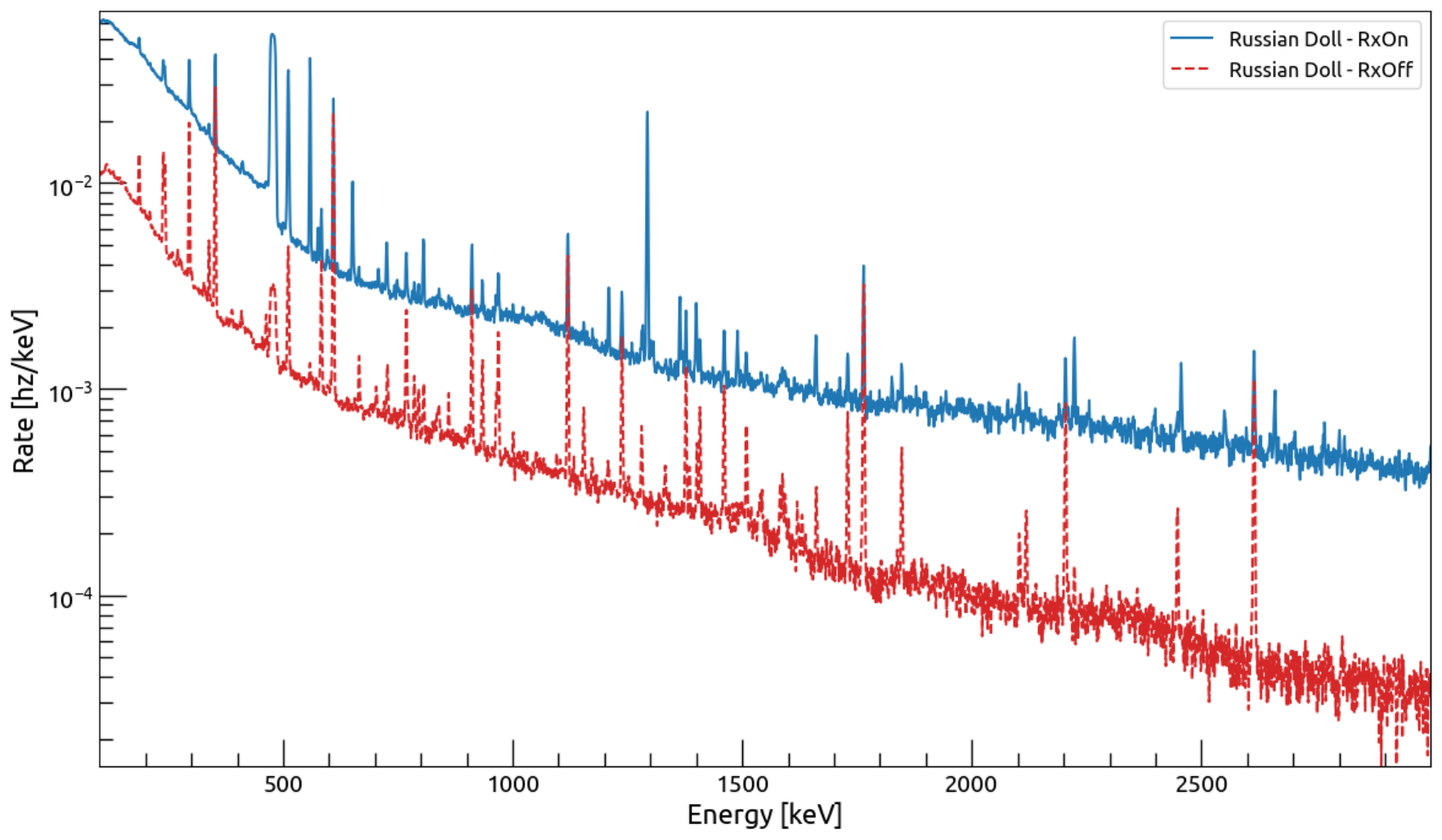}
    \caption{Energy spectrum with Russian doll shielding, red reactor on blue reactor off. Reactor on features are due to neutron captures on shielding materials, especially \el{Cd}{113}.}
    \label{fig:RD_on_off_compare}
\end{figure}

The broad feature centered at 478~keV is due to the Doppler-broadening gamma ray emitted from the \el{Li}{7} nucleus as a result of the $^{10}$B $(n,\alpha)^{7}$ Li reaction, which occurs when neutrons reach the borated polyethylene layer.
As illustrated in Figure~\ref{fig:RD_on_off_compare}, this Doppler broadened feature is also discernible when the reactor is off due to neutrons from cosmogenic backgrounds.
For Table~\ref{tab:rd_lines}, relative peak areas were obtained from background-subtracted Gaussian peak fits to the reactor-on low-gain runs, normalized run-by-run to the \el{Cd}{113} 558.5~keV line, and then combined with inverse-variance weighting. The quoted uncertainties are the propagated fit-area uncertainties carried through that weighted-average procedure.

\begin{table}[pos=htpb]
    \caption{Gamma peaks visible within the Russian doll while reactor is on. "n,g" denotes neutron capture on the listed isotope, "b-" denotes radioactive beta decay of the isotope. Peak areas are relative to the \el{Cd}{113} 558.5 keV peak and are taken from Cycle 498 data (no additional lead / water shield). The radioactive decays are also visible during reactor off. }
    \label{tab:rd_lines}
    \centering
    \begin{tabular}{llll}
\toprule
& energy [keV] & ID & rel. area (Rx ON) \\
\midrule
& 238.6 & \el{Pb}{212} b- & 0.346 $\pm$ 0.049 \\
& 242 & \el{Pb}{214} b- & 0.272 $\pm$ 0.040 \\
& 295.2 & \el{Pb}{214} b- & 0.455 $\pm$ 0.793 \\
& 351.9 & \el{Pb}{214} b- & 0.758 $\pm$ 0.547 \\
& 478 & \el{B}{10}(n,$\alpha$)\el{Li}{7} & - \\
& 558.5 & \el{Cd}{113} n,g & 1.000 $\pm$ 0.193 \\
& 609.3 & \el{Bi}{214} b- & 0.605 $\pm$ 0.143 \\
& 651.3 & \el{Cd}{113} n,g & 0.173 $\pm$ 0.141 \\
& 707.4 & \el{Cd}{113} n,g & 0.018 $\pm$ 0.110 \\
& 725 & \el{Cd}{113} n,g & 0.060 $\pm$ 0.087 \\
& 768.4 & \el{Bi}{214} b- & 0.047 $\pm$ 0.101 \\
& 805.9 & \el{Cd}{113} n,g & 0.106 $\pm$ 0.080 \\
& 1120.3 & \el{Bi}{214} b- & 0.116 $\pm$ 0.058 \\
& 1209.7 & \el{Cd}{113} n,g & 0.050 $\pm$ 0.044 \\
& 1238.1 & \el{Bi}{214} b- & 0.035 $\pm$ 0.056 \\
& 1281 & \el{Bi}{214} b- & 0.024 $\pm$ 0.053 \\
& 1293.6 & \el{Ar}{41} b- & 0.565 $\pm$ 0.087 \\
& 1364.3 & \el{Cd}{113} n,g & 0.060 $\pm$ 0.038 \\
& 1377.7 & \el{Bi}{214} b- & 0.037 $\pm$ 0.042 \\
& 1399.6 & \el{Cd}{113} n, g & 0.035 $\pm$ 0.040 \\
& 1489.6 & \el{Cd}{113} n,g & 0.014 $\pm$ 0.048 \\
& 1660.4 & \el{Cd}{113} n,g & 0.032 $\pm$ 0.029 \\
& 1764.5 & \el{Bi}{214} b- & 0.091 $\pm$ 0.026 \\
& 2204.2 & \el{Bi}{214} b- & 0.021 $\pm$ 0.023 \\
& 2223 & \el{H}{1} n,g & 0.033 $\pm$ 0.020 \\
& 2398.6 & \el{Cd}{113} n,g & 0.008 $\pm$ 0.021 \\
& 2455.8 & \el{Cd}{113} n,g & 0.017 $\pm$ 0.023 \\
& 2550.1 & \el{Cd}{113} n,g & 0.003 $\pm$ 0.023 \\
& 2614.5 & \el{Tl}{208} b- & 0.024 $\pm$ 0.016 \\
& 2660.1 & \el{Cd}{113} n,g & 0.025 $\pm$ 0.012 \\
& 2767.5 & \el{Cd}{113} n,g & 0.011 $\pm$ 0.015 \\
& 5433.1 & \el{Cd}{113} n,g & 0.002 $\pm$ 0.001 \\
& 5824.6 & \el{Cd}{113} n,g & 0.005 $\pm$ 0.001 \\
& 7367.9 & \el{Pb}{207} n,g & 0.003 $\pm$ 0.000 \\
& 7916.3 & \el{Cu}{63} n,g & 0.001 $\pm$ 0.000 \\
\bottomrule
\end{tabular}
\end{table}

\subsection{Shield Configuration Comparison}
\label{sec:rd_shield_compare}
Shielding from thermal neutrons is important to mitigate secondary gammas from capture on shielding materials.
Water is an economical means of such shielding; with a low average mass per nucleon, it is an effective absorber of neutrons.
Measurements were taken with four configurations of shielding. The first used only the layered shielding described in Section~\ref{sec:russian_doll}.
The second added a 7-layer square wall of water bricks around the full RD assembly for a total of 60.5 water bricks (some half bricks of dimension 9$\times$9$\times$6~inch were used).
The third configuration used a 6-layer water-brick wall together with a tiled lead layer placed directly below the cart.
For the final configuration, 8 additional water bricks were added to restore the surrounding wall to 7 layers while keeping the lead-under-cart layer, as pictured in Figure~\ref{fig:rd_water}. A scaled schematic of each configuration is shown in Figure~\ref{fig:water_brick_configs}. The live times were not equal across these shielding configurations; the added-shield runs span from roughly 50~h to roughly 1800~h, so the comparisons below should be interpreted through the normalized rates rather than raw counting totals.

Table~\ref{tab:rd_rates} presents a comparison of gamma rates for various shield configurations, both at low energies (30--60 keV) and across the full energy range (50--11,500 keV). Although runs were not conducted in low-gain mode (11.4 MeV range) for all shielding configurations, rates are provided for the available data in those instances.

The addition of water and lead shielding reduces the RD reactor-only rate substantially, leaving reactor-on rates only about three times the reactor-off levels. This indicates that the measurement is approaching a regime in which ambient radioactivity and cosmogenic backgrounds contribute comparably to the residual reactor-related component.

\begin{figure*}[pos=htbp]
    \centering
    \includegraphics[width=1.0\linewidth]{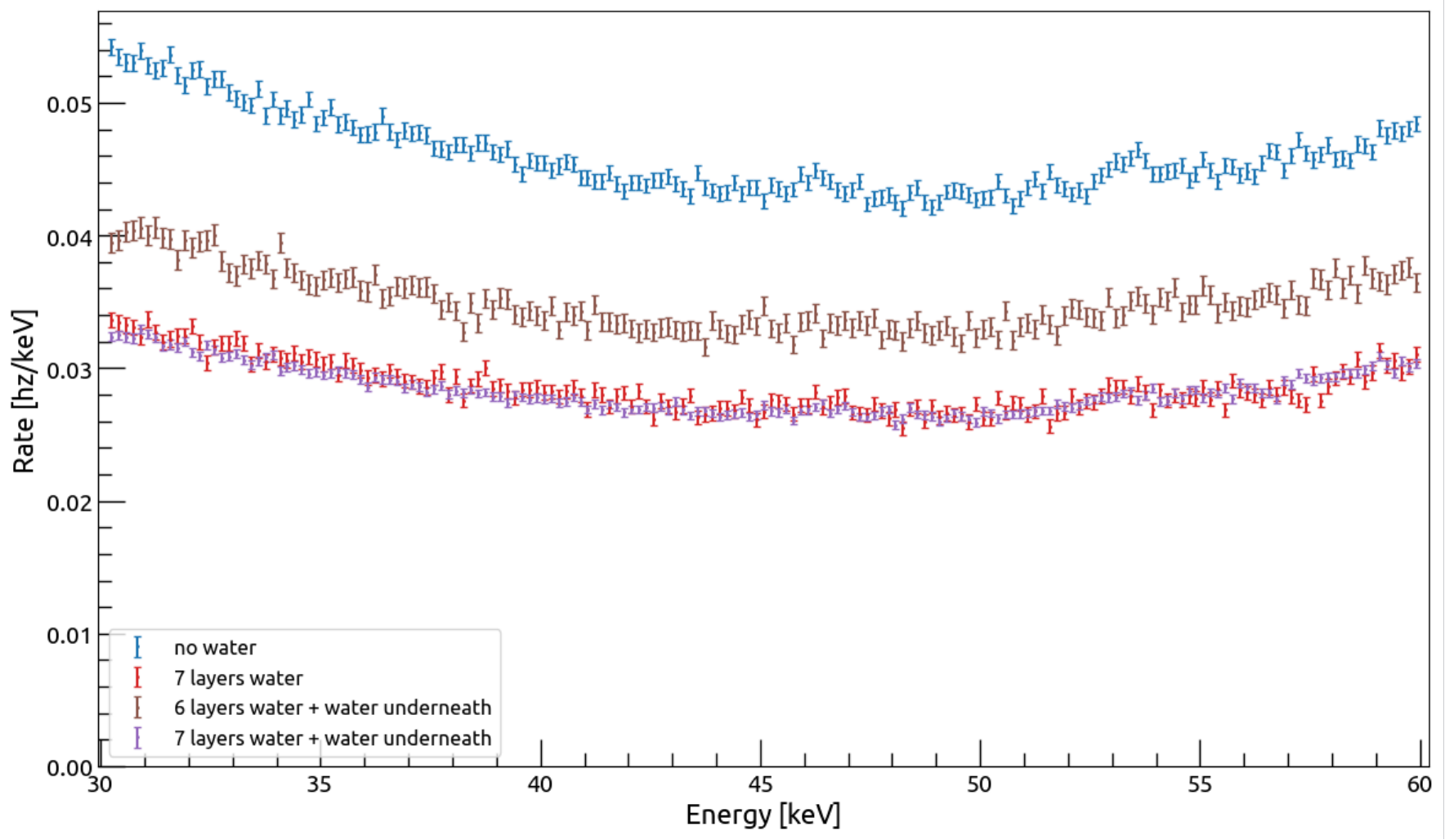}
    \caption{Energy spectra taken in the Russian Doll shield in high gain mode (30--60 keV). Comparison shown is between different shield configurations described in Section~\ref{sec:rd_shield_compare}. Error bars are from counting statistics.}
    \label{fig:rd_shield_comparison}
\end{figure*}

In the low-energy region (30--60 keV), the spectrum exhibits a predominantly flat background, as shown in Figure~\ref{fig:rd_shield_comparison}. This range was selected based on the lower energy threshold of the Ge detector, where the thickness of the outer dead layer limits the minimum detectable energy to 30 keV. Future efforts could involve deploying specialized detectors to assess backgrounds at sub-keV energies, which are more pertinent to~\cevns~detection.

Using the reactor-only rates in Table~\ref{tab:rd_rates}, the 7-layer water-brick wall reduces the 30--60~keV background by about 47\% relative to the no-water configuration. This decrease demonstrates the importance of additional shielding against the reactor-related neutron background. The lead-under-cart configurations do not improve dramatically beyond the water-wall-only case in the current dataset, and the open-top geometry means the result should be interpreted as added lateral moderation around the RD assembly rather than a fully enclosed neutron shield.

\renewcommand\cellalign{ll}
\begin{table*}[pos=!tbp]
    \caption{Rates measured within the RD shield for different surrounding-shield configurations. ``Layers'' refers to the number of water-brick layers stacked from the floor to form the square wall around the RD assembly, and ``+'' denotes the additional tiled lead layer placed directly beneath the cart. The 30--60 keV range is measured with a high gain mode, the 50--11380 keV range with the nominal low gain mode used for the position scan studies. Reactor-on and reactor-off rates are shown separately; reactor-only rates are obtained by subtracting reactor-off from reactor-on.}
    \label{tab:rd_rates}
    \centering

    \begin{tabular}{lllllll}
\toprule
 \makecell{shield \\ config} & \makecell{30--60 keV \\ Rx-on \\ {[mHz/kg/keV]}} & \makecell{30--60 keV \\ Rx-off \\ {[mHz/kg/keV]}} & \makecell{30--60 keV \\ Rx-only \\ {[mHz/kg/keV]}} & \makecell{50--11380 keV \\ Rx-on \\ {[mHz/kg/keV]}} & \makecell{50--11380 keV \\ Rx-off \\ {[mHz/kg/keV]}} & \makecell{50--11380 keV \\ Rx-only \\ {[mHz/kg/keV]}} \\
\midrule
 no water         & 46.1 $\pm$ 2.3 & 9.74 $\pm$ 0.49 & 36.3 $\pm$ 2.4 & 1.67 $\pm$ 0.08 & 0.329 $\pm$ 0.016 & 1.34 $\pm$ 0.09 \\
 7 layers         & 28.6 $\pm$ 1.4 & 9.30 $\pm$ 0.47 & 19.3 $\pm$ 1.5 & 1.04 $\pm$ 0.05 & 0.310 $\pm$ 0.016 & 0.729 $\pm$ 0.053 \\
 6 layers+        & 35.1 $\pm$ 1.8 & 9.28 $\pm$ 0.46 & 25.8 $\pm$ 1.8 & ---        & ---  & ---      \\
 7 layers+        & 28.3 $\pm$ 1.4 & 9.38 $\pm$ 0.47 & 18.9 $\pm$ 1.5 & ---        & ---  & ---      \\
\bottomrule
\end{tabular}
\end{table*}

\section{Detector Response Unfolding and Gamma Flux Results}
\label{sec:detector_response}

To estimate the incident gamma flux from measured spectra and provide quantitative source terms for future experiments, a detector response model was created using Geant4. The unfolded gamma flux spectra are available in tabulated format in the supplemental materials. The supplemental files include unfolded spectra for both response-model limits discussed below, metadata tables, migration-matrix support summaries, and calibrated measured spectra.

Within an event, all positions and energies are recorded for energy depositions within the germanium crystal. A spherical charge distribution model approximates charge collection physics. Charge within overlaps of this sphere with detector dead layers is discarded. Dead layer thicknesses are varied based on efficiency measurement fits.
See Section~\ref{sec:efficiency_tuning} for more details.

Total energy from each deposition after dead layer corrections are summed. Detected energy is approximated by sampling a Gaussian with width proportional to $A + B \times E$, where $E$ is the deposited energy and $A$, $B$ are coefficients fit to measured spectral line widths.
More details on determining the energy resolution across the full energy range are found in Section~\ref{sec:photopeak_resolution}.

Detector non-linearity effects, such as those arising from charge trapping at the energy deposition level~\cite{Goulding1987,Arnquist2023}, which can introduce tailing in the pulse shapes particularly for higher-energy events, were not incorporated into the model. This simplification was deemed appropriate, as a precise reproduction of the exact spectral shape was not required for the primary goal of estimating incident gamma flux to understand background contributions in future experiments.

The simulated detector response is calculated for monochromatic incident gamma rays with energies \(E_i\) ranging from 40 to 12,000~keV in 1~keV increments. The observed spectrum \(\mathbf{O}\) (a vector whose elements correspond to binned deposited energies in the detector) is related to the incident spectrum through the equation:
\begin{equation}
\mathbf{O} = \sum_{i} w_i \mathbf{S}_i,
\end{equation}
where \(w_i\) represents the flux weight for the incident gamma-ray energy \(E_i\), and \(\mathbf{S}_i\) is the simulated detector response vector (i.e., the spectrum of deposited energies) for a unit flux of monochromatic gamma rays at incident energy \(E_i\). In matrix notation, this can be expressed as \(\mathbf{O} = \mathbf{M} \mathbf{w}\), where \(\mathbf{M}\) is the response matrix with columns \(\mathbf{S}_i\), and \(\mathbf{w}\) is the vector of flux weights. The Richardson-Lucy algorithm~\cite{richardson,lucy} was employed to perform this deconvolution and recover the incident spectrum.

The gamma field around the lead collimator is not expected to be isotropic, especially at the reactor-facing and beamline-facing positions. At the same time, the measurement campaign does not contain enough directional information to reconstruct the full angular distribution. For that reason, we use two response-model limits: an isotropic field over the outer surface of the collimator-detector system, and a uniform field directed into the plane of the detector-collimator front face. These two cases give a practical bracket on the unfolded flux normalization. Both response matrices now have direct simulated support at 166 energies from 40~keV to 12~MeV.

An isotropic flux of \(f = 1~\mathrm{Hz/mm}^2\) was used in the simulations for the case reported here, such that the deconvolved incident flux spectrum \(\Phi\) (with elements \(\Phi_i\) corresponding to each energy bin \(E_i\)) is given by
\begin{equation}
\Phi_i = w_i \cdot f.
\end{equation}

%The simulated energy spectrum is calculated for energies in 1 keV increments from 30 to 11500 keV by sending monochromatic gamma rays into the detector. The observed spectrum $\mathbf{O}$ (a vector of detected energy bins) is related to the incident spectrum through:

%\begin{equation}
%   \mathbf{O} = \sum_{i} w_i \mathbf{S}_i,
%\end{equation}

%where $w_i$ is the flux weight for incident gamma energy $E_i$, and $\mathbf{S}_i$ is the simulated detector response vector for a unit flux of gammas at energy $E_i$. The Richardson-Lucy algorithm~\cite{richardson,lucy} was used to solve this deconvolution.

%A 1 Hz/mm$^2$ isotropic flux is used in the simulation so that the deconvolved spectrum will consist of the sum of the weights for each energy bin multiplied by this flux.

\subsection{Efficiency Measurements}
\label{sec:efficiency_tuning}
A series of efficiency measurements were performed using the bare Ge detector. A radioactive source was placed at three different distances from the top face of the detector along the cylindrical axis of symmetry.
Photopeak efficiency was measured for 13 different energies by using the sources as summarized in table~\ref{table:eff}.
The setup was simulated in Geant and the dead layers and charge radius parameter as
described in Section~\ref{sec:detector_response} were varied to best match the efficiency data.
The optimal values found were a top dead layer thickness of 1.3 mm, an inner dead layer thickness of 3.6 mm, a side dead layer thickness of 0.65 mm, a bottom dead layer thickness of 3.5 mm, a charge radius parameter of 0.22 mm/MeV$^{1/3}$.

\begin{table*}[pos=!tbp]
    \caption{Efficiency measurement results and simulated values. E = energy, D = distance, Eff = efficiency fractions, meas = measurement, sim = simulation.}
    \label{table:eff}
    \centering
    \begin{tabular}{llllllll}
        \toprule
         & source              & E [keV] & D [mm] & Eff (meas) & Eff (sim) & \% diff & \% meas err \\
        \midrule
         & $^{241}\textrm{Am}$ & 59.5    & 103.7  & 0.00385    & 0.00392   & 1.85    & 5.01        \\
         & $^{241}\textrm{Am}$ & 59.5    & 203.7  & 0.00118    & 0.00119   & 1.32    & 5.03        \\
         & $^{241}\textrm{Am}$ & 59.5    & 403.7  & 0.000317   & 0.000323  & 1.90    & 5.01        \\
         & $^{109}\textrm{Cd}$ & 88      & 103.7  & 0.00832    & 0.00660   & -20.68  & 4.26        \\
         & $^{109}\textrm{Cd}$ & 88      & 203.7  & 0.00249    & 0.00200   & -19.70  & 4.24        \\
         & $^{109}\textrm{Cd}$ & 88      & 403.7  & 0.000693   & 0.000537  & -22.55  & 4.29        \\
         & $^{57}\textrm{Co}$  & 122.1   & 103.7  & 0.00974    & 0.00947   & -2.82   & 3.01        \\
         & $^{57}\textrm{Co}$  & 122.1   & 203.7  & 0.00292    & 0.00288   & -1.51   & 3.02        \\
         & $^{57}\textrm{Co}$  & 122.1   & 403.7  & 0.000814   & 0.000793  & -2.57   & 3.01        \\
         & $^{57}\textrm{Co}$  & 136.5   & 103.7  & 0.00979    & 0.00979   & -0.08   & 3.47        \\
         & $^{57}\textrm{Co}$  & 136.5   & 203.7  & 0.00299    & 0.00300   & 0.42    & 3.51        \\
         & $^{57}\textrm{Co}$  & 136.5   & 403.7  & 0.000826   & 0.000836  & 1.21    & 3.48        \\
         & $^{133}\textrm{Ba}$ & 276     & 103.7  & 0.00724    & 0.00750   & 3.53    & 6.56        \\
         & $^{133}\textrm{Ba}$ & 276     & 203.7  & 0.0023     & 0.00239   & 3.79    & 6.59        \\
         & $^{133}\textrm{Ba}$ & 276     & 403.7  & 0.000655   & 0.000681  & 3.96    & 6.59        \\
         & $^{133}\textrm{Ba}$ & 302.9   & 103.7  & 0.00667    & 0.00658   & -1.32   & 5.89        \\
         & $^{133}\textrm{Ba}$ & 302.9   & 203.7  & 0.00213    & 0.00212   & -0.64   & 5.90        \\
         & $^{133}\textrm{Ba}$ & 302.9   & 403.7  & 0.000609   & 0.000603  & -0.99   & 5.90        \\
         & $^{133}\textrm{Ba}$ & 356     & 103.7  & 0.00595    & 0.00616   & 3.47    & 5.83        \\
         & $^{133}\textrm{Ba}$ & 356     & 203.7  & 0.00191    & 0.00197   & 3.38    & 5.84        \\
         & $^{133}\textrm{Ba}$ & 356     & 403.7  & 0.000548   & 0.000565  & 3.08    & 5.84        \\
         & $^{133}\textrm{Ba}$ & 384     & 103.7  & 0.00563    & 0.00550   & -2.35   & 5.51        \\
         & $^{133}\textrm{Ba}$ & 384     & 203.7  & 0.0018     & 0.00176   & -2.38   & 5.52        \\
         & $^{133}\textrm{Ba}$ & 384     & 403.7  & 0.000519   & 0.000502  & -3.34   & 5.52        \\
         & $^{137}\textrm{Cs}$ & 661.7   & 103.7  & 0.00368    & 0.00371   & 0.80    & 3.02        \\
         & $^{137}\textrm{Cs}$ & 661.7   & 203.7  & 0.0012     & 0.00120   & 0.39    & 3.03        \\
         & $^{137}\textrm{Cs}$ & 661.7   & 403.7  & 0.000349   & 0.000349  & -0.12   & 3.04        \\
         & $^{54}\textrm{Mn}$  & 834.8   & 103.7  & 0.00311    & 0.00311   & 0.11    & 3.01        \\
         & $^{54}\textrm{Mn}$  & 834.8   & 203.7  & 0.00102    & 0.00102   & -0.05   & 3.02        \\
         & $^{54}\textrm{Mn}$  & 834.8   & 403.7  & 0.000296   & 0.000296  & -0.07   & 3.05        \\
         & $^{60}\textrm{Co}$  & 1173.2 & 103.7  & 0.00238    & 0.00241   & 1.05    & 3.01        \\
         & $^{60}\textrm{Co}$  & 1173.2 & 203.7  & 0.000798   & 0.000795  & -0.35   & 3.03        \\
         & $^{60}\textrm{Co}$  & 1173.2 & 403.7  & 0.000233   & 0.000232  & -0.28   & 3.07        \\
         & $^{22}\textrm{Na}$  & 1275    & 103.7  & 0.00221    & 0.00227   & 2.69    & 3.01        \\
         & $^{22}\textrm{Na}$  & 1275    & 203.7  & 0.000747   & 0.000753  & 0.78    & 3.02        \\
         & $^{22}\textrm{Na}$  & 1275    & 403.7  & 0.000221   & 0.000220  & -0.49   & 3.03        \\
         & $^{60}\textrm{Co}$  & 1332.5  & 103.7  & 0.00217    & 0.00219   & 0.72    & 3.01        \\
         & $^{60}\textrm{Co}$  & 1332.5  & 203.7  & -          & 0.000727  & -       & -           \\
         & $^{60}\textrm{Co}$  & 1332.5  & 403.7  & 0.000215   & 0.000213  & -0.71   & 3.06        \\
        \bottomrule
    \end{tabular}
\end{table*}

A lack of efficiency data at high energy means we have to rely on the accuracy of the Geant simulation for reconstructing the gamma flux.
An experiment to determine the efficiency at higher energies is difficult due to the issue of finding a source in the 7+ MeV range with a well known activity.
We can instead use the high energy photopeaks from neutron capture sources in the reactor to measure the relative intensity of the full absorption peak and the first and second escape peaks to validate the simulation at high energies. This is shown in Appendix~\ref{app:validation}.

\subsection{Photopeak Resolution}
\label{sec:photopeak_resolution}
Spectra taken when the detector was pointing towards the reactor at the MIF location were ideal for measuring the detector resolution at the full range of energies. The 11.4 MeV peak from $^{59}\textrm{Ni}$ provides a near background-free measure of the resolution at the upper end of the energy range. Additionally, the 9.0 MeV peak from \el{Ni}{58} and the 7.7 MeV peak from $^{27}\textrm{Al}$ are well above background.

This data was also useful for validating the simulation at high energy. The simulation should be able to accurately determine the relative intensities of the full absorption peak, the first escape peak, and the second escape peak for a given energy. This is shown in Appendix~\ref{app:validation}.

Peaks were fit using a Gaussian plus a skew Gaussian plus a quadratic background. The functional form of the fit $f_i$ for the $i^{\rm th}$ peak is

\begin{equation}
    \label{eq:peak_fit}
    \begin{split}
        f_i(E) = & H_i (1 - R_i) \, \mathrm{exp}\left[- \frac{(E - c_i)^{2}}{2\sigma_i^2}\right] \\
                 & + H_i  R_i \,  \mathrm{exp}\left[\frac{E - c_i}{\beta_i}\right]  \mathrm{erfc}\left[
            \frac{E - c_i}{\sqrt{2}\sigma_i} + \frac{\sigma_i}{\sqrt{2}  \beta_i}\right] \\
                 & + D_i + F_i E + G_i E^2
    \end{split}
\end{equation}
where $E$ is the energy, $H_i$ is the amplitude, $R_i$ is the fraction of the amplitude attributed to the skew Gaussian component, $c_i$ is the centroid, $\sigma_i$ is the standard deviation, $\beta_i$ is the skewness parameter, $\mathrm{erfc}$ is the complementary error function, and $D_i$, $F_i$, and $G_i$ are coefficients for the quadratic background.

The skew Gaussian accounts for charge loss due to impurities in the crystal while the Gaussian accounts for variations in the amount of charge collected due to thermal fluctuations and electronic noise.

\begin{figure}[pos=htbp]
    \centering
    \includegraphics[width=1.0\linewidth]{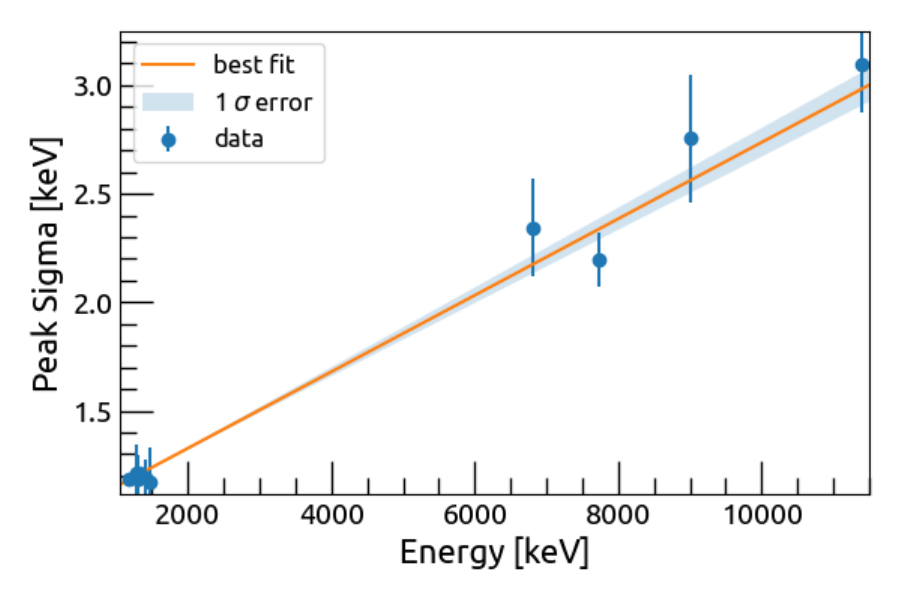}
    \caption{Linear fit to sigma widths as a function of energy. Fits were performed on the major gamma lines from $^{59}\textrm{Ni}$, $^{58}\textrm{Ni}$, $^{27}\textrm{Al}$, $^{9}\textrm{Be}$, $^{41}\textrm{Ar}$, $^{40}\textrm{K}$, and $^{60}\textrm{Co}$.}
    \label{fig:sigma_fit}
\end{figure}

A linear function $A + B \times E$ was used to fit the sigma widths of the peaks as a function of energy. The fit values were
$A$ = 0.98 keV, $B$ = 0.00018. The linear fit is displayed in Figure~\ref{fig:sigma_fit}, and results of each peak fit along with their individual \chisqr~values are displayed in table~\ref{tab:peak-fit}. 

\begin{table}[pos=htbp]
\centering
\caption{Peak fit results}
\label{tab:peak-fit}
\begin{tabular}{ccccc} 
\toprule
Energy [keV] & $\sigma$ [keV] & $\Delta \sigma$ [keV] & $\chi^2$ & ndf \\ 
\midrule
1173.20 & 1.18 & 0.01 & 1.98 & 5 \\ 
1274.40 & 1.21 & 0.13 & 7.14 & 5 \\ 
1293.64 & 1.18 & 0.11 & 2.06 & 5 \\ 
1332.50 & 1.21 & 0.01 & 0.81 & 5 \\ 
1408.01 & 1.19 & 0.08 & 3.46 & 5 \\ 
1460.80 & 1.17 & 0.15 & 2.44 & 5 \\ 
6809.61 & 2.34 & 0.22 & 4.73 & 7 \\ 
7724.03 & 2.20 & 0.13 & 6.63 & 9 \\ 
8998.63 & 2.76 & 0.29 & 7.24 & 9 \\ 
11386.50 & 3.09 & 0.22 & 43.03 & 25 \\ 
\bottomrule
\end{tabular}
\end{table}

\subsection{Simulation Validation}
\label{sec:simulation_validation}
The simulation was validated at high energies by comparing measured ratios of the full absorption peak to the first and second escape peaks for selected neutron capture gamma lines observed in reactor spectra. Detailed results, including comparisons between data and simulations for the isotropic and front-face response-model limits, are presented in Appendix~\ref{app:validation}, Table~\ref{table:peak_ratio_compare}. In general, the simulated ratios agree well with measurements, particularly for the isotropic case, confirming the accuracy of the geometry, Compton scattering, photoelectric effect, and pair production physics in the model. This comparison also supports the limiting-case treatment described above, because the data indicate a substantial contribution from gammas entering through lead-shielded directions.
During this validation process, we also compared the two response-model limits used in the unfolding: an isotropic field over the outer surface of the collimator-detector system and a uniform field directed into the detector-collimator front face. The detailed results are shown in Appendix~\ref{app:validation}, Figure~\ref{fig:collimator_effectiveness}. The front-face assumption yields a much larger source-area-normalized photopeak rate than the isotropic assumption, by about a factor of 120 at 60--80~keV and about 15 at 2000~keV. This large low-energy separation and more modest high-energy separation help explain why the cart pointing information is most useful at low energies and provides only limited discrimination once multi-MeV gammas can penetrate the lead from many directions.

\subsection{Unfolded Gamma Flux Spectra at Key Locations}
\label{sec:unfolded_spectra}

\begin{figure}[pos=htbp]
\centering
\includegraphics[width=0.95\linewidth]{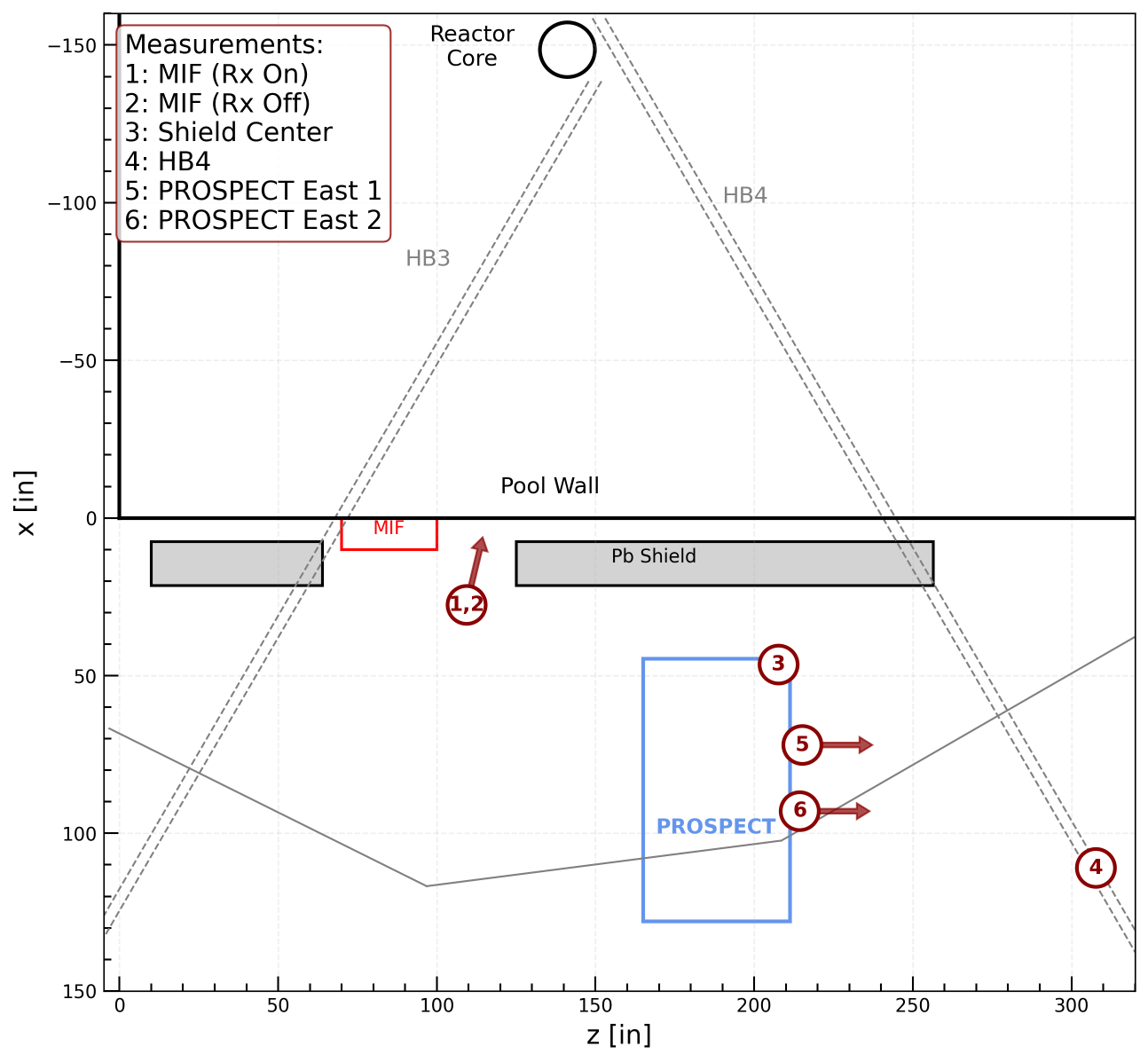}
\caption{Top-down view of the HFIR reactor area showing all measurement locations analyzed in this study. Each numbered circle indicates a measurement position. Arrows indicate the pointing direction for collimated measurements where the detector was not oriented vertically downward. The PROSPECT detector position, reactor core, beam lines (HB3 and HB4), lead shielding walls, and pool boundary are shown for reference.}
\label{fig:measurement_locations}
\end{figure}

The Richardson-Lucy unfolding procedure described in Section~\ref{sec:detector_response} was applied to measurements at six locations in the HFIR reactor pool area, with detector orientations ranging from downward-facing (vertical) to horizontal toward specific background sources. Figure~\ref{fig:measurement_locations} shows the spatial distribution of these positions, numbered sequentially for reference.
Each location probes a different mix of background sources and shielding conditions. The unfolded spectra give the incident gamma flux as a function of energy, which makes it possible to compare locations and to build source terms for transport studies. The supplemental CSV files include both the isotropic and front-face limiting-case unfolds, along with the calibrated measured spectra and metadata for all six positions shown in Figure~\ref{fig:measurement_locations}. The file format is described in the supplemental README.

Figure~\ref{fig:unfold_comparison_example} shows a side-by-side comparison of the raw measured spectra and the unfolded incident gamma flux spectra for all six measurement locations. The dual-axis presentation illustrates how the unfolding procedure transforms the detector response into estimated incident flux, removing detector effects such as Compton scattering, escape peaks, and energy-dependent efficiency.

\begin{figure*}[pos=htbp]
\centering
\includegraphics[width=1.0\textwidth]{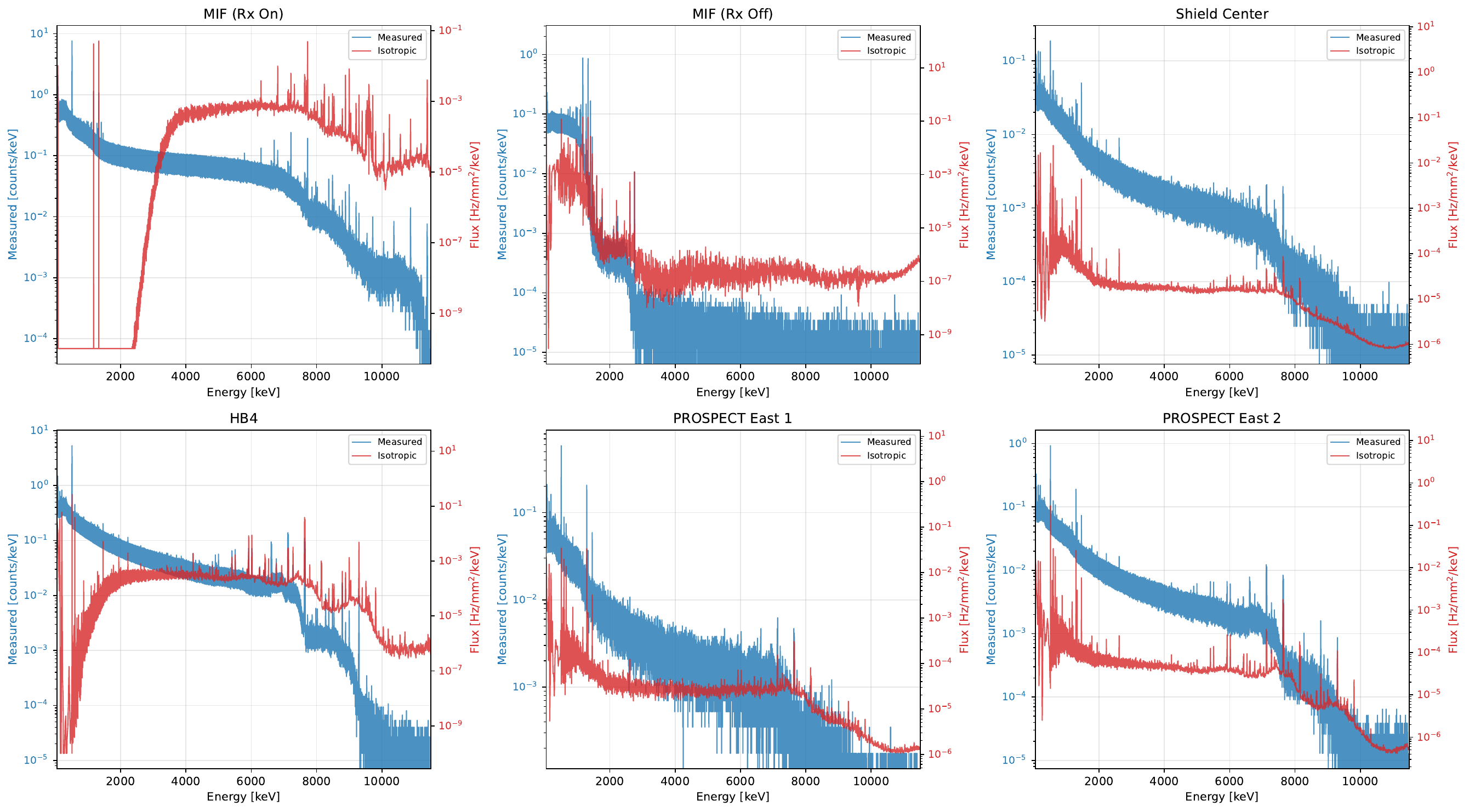}
\caption{Comparison of raw measured spectra (blue, left axis) and unfolded gamma flux spectra (red, right axis) for each measurement location. The measured spectra represent the detector response while the unfolded spectra represent the estimated incident gamma flux.}
\label{fig:unfold_comparison_example}
\end{figure*}

\begin{figure*}[pos=tbp]
\centering
\includegraphics[width=\textwidth]{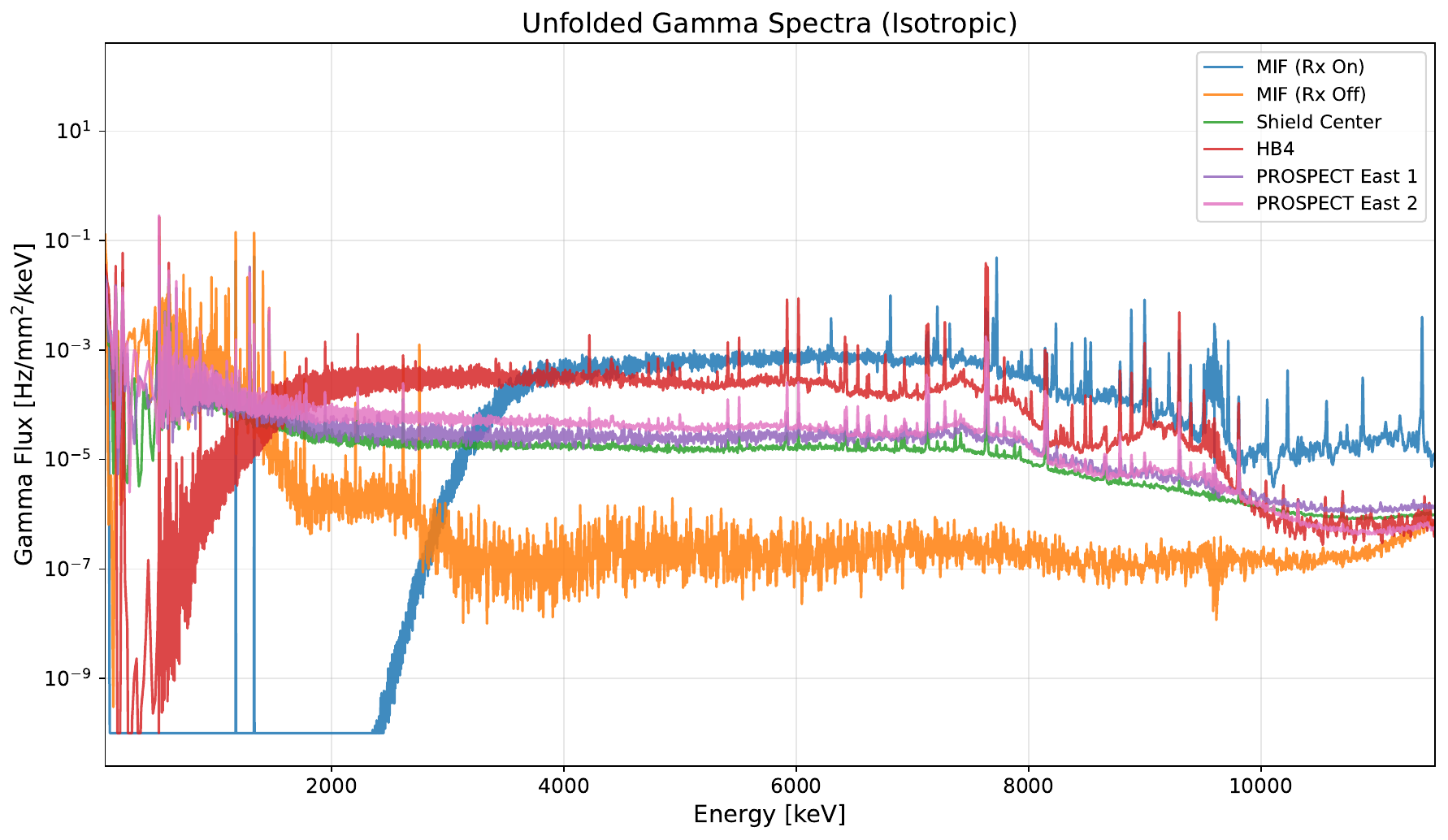}
\caption{Unfolded incident gamma flux spectra for all six measurement locations using the isotropic-response case. Vertical units are Hz/mm$^2$/keV. Measurement numbers correspond to the positions in Figure~\ref{fig:measurement_locations}.}
\label{fig:all_unfolded_spectra}
\end{figure*}

The unfolded incident gamma flux spectra for all six locations are shown in Figure~\ref{fig:all_unfolded_spectra} using the isotropic-response assumption. The spectra vary strongly by location, shielding setup, and source geometry.
At five of the six locations, the front-face limiting case stays below the isotropic response through most of the 2--9~MeV band. The reactor-facing MIF point is the most obvious exception: there the curves stay close together, and the front-face limit raises the 0.5--2~MeV flux where the isotropic result shows a dip. That pattern is most consistent with a local geometry mismatch at that one position, not a broader failure of the detector or collimator model. For that location, the front-face limit looks like the more realistic illumination than the isotropic assumption.

The isotropic assumption can produce unphysical features at locations with highly directional flux, such as the MIF reactor-facing position. In particular, the isotropic model attributes too many high-energy gammas to directions that would not contribute in a directional field, and the resulting excess of downscattered events at lower energies absorbs the flux budget that should be assigned to actual low-energy gammas. This artifact manifests as a suppression of the unfolded flux below $\sim$2~MeV for the MIF reactor-on case and similarly explains why the HB4 unfolded flux can appear lower than East~1 or East~2 at low energies despite having higher measured count rates. The front-face response model provides a directional limiting case that bounds these features in the finalized response-model bracket.

To check whether the MIF low-energy suppression is specific to the Richardson-Lucy procedure, we repeated the unfolding with an independent non-negative solver that minimizes a penalized Poisson objective using mirror-descent updates~\cite{beck_teboulle_mirror,bardsley_poisson_tikhonov}. For the reactor-facing MIF spectrum, the suppression remains when the isotropic response is used, while the front-face result changes very little. In the nominal comparison, the 0.05--2~MeV fraction for MIF changes from 0.0611 to 0.0572 under the isotropic response and from 0.1063 to 0.1068 under the front-face response when Richardson-Lucy is replaced by the alternate solver. We also propagated counting-statistics fluctuations from the measured Ge spectrum through matched toy unfolds. For MIF reactor-on, the algorithm-to-algorithm shift in the 0.05--2~MeV fraction is $-0.0032 \pm 0.0014$ for the isotropic response and $+0.0004 \pm 0.0002$ for the front-face response, whereas the front-minus-isotropic shift is $+0.0467 \pm 0.0029$ for Richardson-Lucy and $+0.0503 \pm 0.0038$ for the alternate solver. The persistence of the low-energy suppression across unfolding algorithms therefore indicates that the dominant uncertainty comes from the detector-response simulation used to construct the migration matrix, rather than from the Richardson-Lucy method itself. Changing the assumed illumination geometry modifies the size of the effect, but does not eliminate it, which points to incomplete modeling of the detector, collimator, and surrounding shielding response rather than to a purely algorithmic artifact.

The HB4 location provides a useful second directional check. There, the front-face unfold is again nearly unchanged across the two solvers, but the isotropic solution is more solver-sensitive than it is for MIF. We therefore interpret the front-face stability as the more robust statement and do not treat the detailed low-energy shape of the isotropic HB4 result as algorithm-independent.

\begin{figure*}[pos=htbp]
\centering
\includegraphics[width=1.0\linewidth]{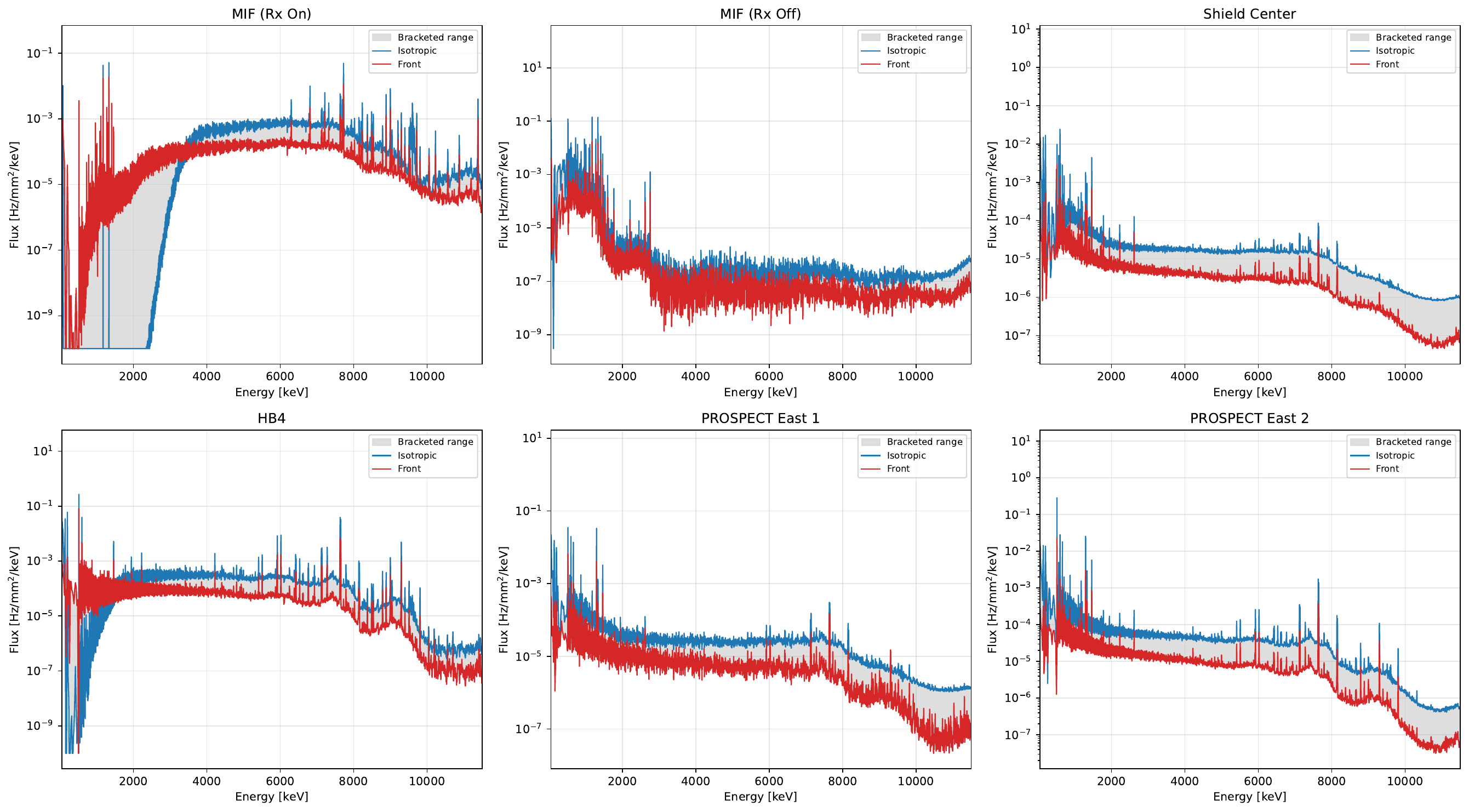}
\caption{Comparison of the two unfolding response-model limits for each measurement location. Blue curves show the isotropic response, red curves show the front-face directional response, and the shaded band indicates the bracket between them. This band is used as the plausible response-model range for the unfolded incident gamma flux.}
\label{fig:unfolded_spectrum_bounds}
\end{figure*}

Figure~\ref{fig:unfolded_spectrum_bounds} compares the isotropic and front-face unfolds directly. The shaded region is the current response-model bracket on the incident flux. For most locations, the front-face response pulls the mid-energy continuum below the isotropic result. At the reactor-facing MIF position, the main difference is instead at lower energy, where the front-face response removes the unphysical 0--2~MeV dip seen in the isotropic unfold. Above about 9~MeV, the two response-model cases often move back toward each other, which is reasonable if the highest-energy gammas are less sensitive to the assumed angular distribution after passing through the shielding.

The unfolded spectra provide source terms for radiation transport simulations. The supplemental CSV files include both response-model limits over the unfolded energy range used in this work, along with the measured spectra and metadata needed to track which limiting case is being used.

\section{Gamma Flux Model Consistency Check and Calibration Studies with \pspt~Data}
\label{sec:prospect_gammas}

The primary rationale for this Section is to check consistency of the unfolded gamma source term models obtained in Section~\ref{sec:detector_response} with \pspt~data. An additional motivation is to explore whether features in the gamma background at HFIR can be used to aid energy calibration of future experiments.

Source calibrations with the \pspt~detector rely on internal calibration tubes which allow radioactive source capsules to be placed in between segments at any position along their length~\cite{PROSPECT}.
Future experiments may not utilize internal calibration tubes, opting for a simplified enclosure for the liquid scintillator volume~\cite{P2_calibration}.
This means that energy calibrations would have to be done with sources external to the detector.

In this Section we explore whether there are features in the gamma background at HFIR that can be used to aid the energy calibration of a future experiment.

\subsection{Gamma Measurements at \pspt~Boundary}
Gamma measurements were made at the boundary of the \pspt~inner scintillator volume in the absence of the PROSPECT detector and its associated shielding package as described in Section~\ref{sec:position_scan}.
That Section showed how the rates throughout the region of the \pspt~detector varied, as seen in Figures~\ref{fig:down_scan} and ~\ref{fig:east_scan}.
A comparison of energy spectra taken with the collimated Ge detector at two locations along the \pspt~boundary is displayed in Figure~\ref{fig:prospect_boundary_gamma}.
This shows that not only are the background rates changing as a function of position, but so is the relative energy spectrum.
This is due to the varying sources of background within the vicinity.

From studying the spectra at the MIF location and the HB4 hotspot (see Figure~\ref{fig:HB4_shield} for the gamma spectra), we see that gammas coming from HB4 do not have lines from neutron capture on Be, Cr, Cu, or \el{Ni}{59}.
Conversely, no Ti lines are seen in the spectra coming from the reactor.
Additionally, the Al and Fe lines which are dominant will be present in varied amounts depending on the location due to the varying amounts and compositions of structural material throughout the building.

For these reasons, the gammas present throughout the \pspt~volume will vary in energy composition.
This makes it challenging to precisely simulate the features of the \pspt~gamma spectrum.
However, the high statistics spectrum taken at the MIF location shows that the \el{Ni}{59}(n,$\gamma$)\el{Ni}{60}~line at 11.39 MeV is the only gamma line visible above 10.5 MeV.
The lack of reactor-related backgrounds near 11.4 MeV makes the \el{Ni}{59} neutron capture line an ideal candidate for an energy calibration point well above those used by PROSPECT~\cite{PROSPECT} and other reactor experiments~\cite{STEREO_calibration}.
In the next Sections we will briefly describe the \pspt~detector, data analysis, and the \pspt~simulation package, used to validate the unfolded gamma flux and check the viability of using the 11.39 MeV gamma to aid in calibration.

\subsection{PROSPECT Detector and Data Analysis}
\label{sec:prospect_detector_analysis}

The PROSPECT detector is a segmented, reactor-based antineutrino spectrometer designed for short-baseline oscillation measurements and precise spectral characterization. It consists of a 4-ton active volume of $^6$Li-doped liquid scintillator (LiLS), divided into an 11$\times$14 array of optically isolated segments, each with dimensions of approximately 117.6 cm in length and 14.5 cm $\times$ 14.5 cm in cross-section. This segmentation enables enhanced position reconstruction, background rejection, and topological discrimination of inverse beta decay (IBD) events through the spatial correlation of prompt positron and delayed neutron capture signals. The segments are housed within an acrylic vessel, surrounded by reflective panels and support structures, and read out by dual photomultiplier tubes (PMTs) at each end to provide energy and timing information.

Data acquisition in PROSPECT employs a zero-suppressed waveform digitization scheme, where only signals exceeding a predefined threshold are recorded to optimize storage and processing efficiency. These waveforms are subsequently reconstructed into individual pulses via baseline subtraction, integration, and timing extraction algorithms. Pulses are then grouped into clusters based on time proximity, forming event candidates that represent physical interactions within the detector.

\begin{figure*}[pos=htbp]
    \centering
    \includegraphics[width=1.0\linewidth]{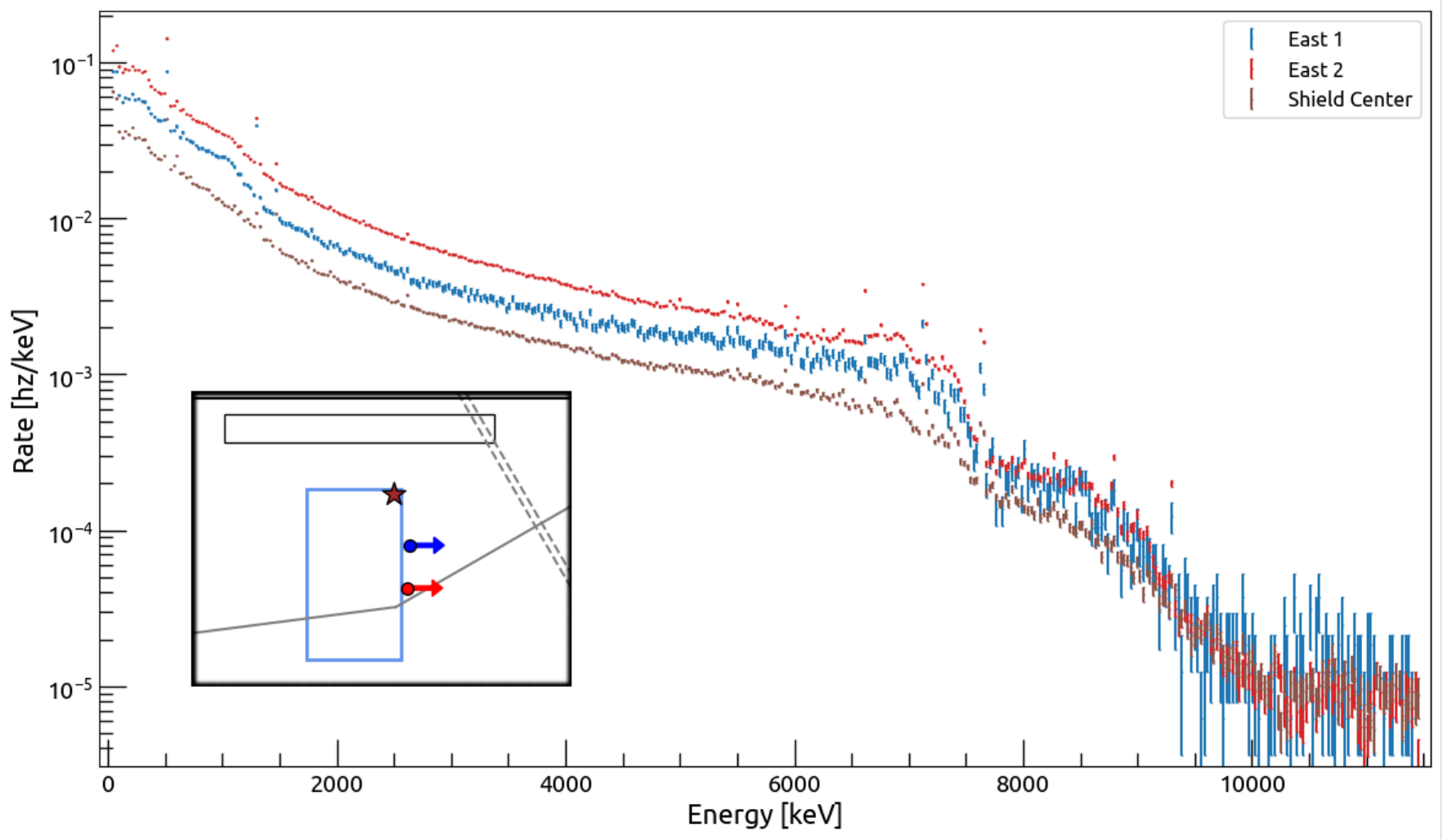}
\caption{Comparison of overnight spectra acquired with the collimated Ge detector at three locations along the \pspt~detector boundary. Inset depicts each measurement's position relative to the PROSPECT scintillator volume boundary (light blue), with the solid line above the points representing the reactor pool wall, the grey rectangle the lead shield wall, upper-right dotted lines the HB4 beamline, and grey lines beneath the detector region the monolith boundary. The blue arrow indicates the `East' location for the blue spectrum, with the arrow denoting collimator facing direction; similarly the red arrow shows the `East 2' orientation and the brown star denotes the `Shield Center' location, with the collimator facing directly down. The plotted range emphasizes the part of the spectrum where location-dependent continuum differences are most visually apparent; the isolated 11.39~MeV \el{Ni}{59} feature is discussed separately below.}
    \label{fig:prospect_boundary_gamma}
\end{figure*}

The analysis presented herein utilizes gamma-ray data collected at the Shield Center location, as depicted in Figure~\ref{fig:prospect_boundary_gamma}. This dataset is selected due to its proximity to the reactor core, capturing contributions from high-energy gammas, including those from neutron capture on $^{59}$Ni.

To obtain a spectrum of reactor-gamma-like events, as shown in Figure~\ref{fig:prospect_nw_sim}, a series of analysis cuts are applied. Pulse-shape discrimination is employed to distinguish gamma-like interactions from other particle types. In addition, background subtraction is performed using an on-off technique, where spectra from reactor-on periods are compared to those from reactor-off periods to subtract non-reactor-related contributions, such as cosmogenic and environmental backgrounds. More details on pulse-shape discrimination and background subtraction can be found in~\cite{PROSPECT}.

\begin{figure}[pos=htbp]
    \centering
    \includegraphics[width=1.0\linewidth]{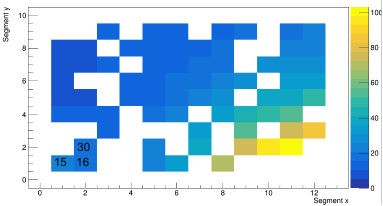}
    \caption{Prompt-like detected singles rates (Hz) during reactor-on vs segment number. Segments 15, 16, and 30 used for gamma simulation comparison are labeled. Missing segments had non-functioning PMTs on one or both ends during the measurement campaign. Source:~\cite{PROSPECT}}
    
    \label{fig:prospect_IBD_efficiency}
\end{figure}

The long axis of each scintillator segment runs east to west. Segment labeling follows a coordinate system where rows are indexed from bottom to top (e.g., row 0 is closest to the floor of the experimental hall) and columns from north to south (along the $x$ axis), as seen in Figure~\ref{fig:prospect_IBD_efficiency}. This plot shows signal rates for gamma-like events taken during reactor-on, which shows how the HB4 hot spot increase detected rates for the segments closer to the floor and farther south (larger $x$ values). 

\subsection{PROSPECT Simulation and Analysis Description}
\label{sec:PG4}
The PROSPECT experiment employs a Monte Carlo simulation framework to model the detector's response to various particles and radiation sources~\cite{PROSPECT}. This simulation is essential for relating the antineutrino energy to the visible energy deposited by positrons in inverse beta decay (IBD) interactions, as well as for validating calibration procedures and background estimations.

The core simulation package utilized is PG4, which is based on the Geant4 toolkit. PG4 incorporates a detailed geometric model of the PROSPECT detector, including the lithium-loaded liquid scintillator (LiLS) volume, acrylic and steel containers, photomultiplier tube (PMT) housings and openings, reflective panels, support rods, calibration tubes, source enclosures, and surrounding shielding structures. This comprehensive geometry ensures accurate accounting of energy losses due to interactions with non-active materials and edge effects as a function of particle energy.

In PG4, Geant4 handles the transport of particles through the detector, tracking energy depositions within the scintillator. These depositions are recorded in an HDF5 file for subsequent analysis. To optimize computational efficiency, the simulation does not explicitly model the scintillation process or the propagation of optical photons. Instead, a parameterized approach is adopted to estimate light yield and quenching effects.

Particle tracks are discretized into steps by Geant4, with step lengths determined by the particle's energy and interaction probabilities. At each step, the energy deposited via ionization is computed, and the corresponding scintillation light output is calculated using Birks' empirical quenching model~\cite{Birks:1964zz}:

\begin{equation}
    \label{eq:birks}
    \frac{dE_{scint}}{dx} = \frac{\frac{dE}{dx}}{1 + k_{B}\frac{dE}{dx}},
\end{equation}

where \(dE/dx\) is the ionization energy loss per unit length, and \(k_B\) is the first-order Birks' constant, fitted to experimental data.

For high-energy electrons from gamma or cosmic interactions, Cherenkov light can also contribute to the visible signal. In PG4, this contribution is modeled phenomenologically as an additional light-yield term proportional to the calculated Cherenkov photon production:

\begin{equation}
    \label{eq:cherenkov_pg4}
    E_c = k_c \sum_{\lambda} N_{\lambda} E_{\lambda},
\end{equation}

where \(E_c\) represents the effective visible-energy contribution associated with Cherenkov photon production and collection, \(k_c\) is a normalization parameter tuned to the scintillator's refractive index, \(N_{\lambda}\) is the number of photons at wavelength \(\lambda\), and \(E_{\lambda}\) is their energy.

The total scintillation energy for a particle track is the sum over all steps of contributions from both ionization and Cherenkov processes:

\begin{equation}
\label{eq:energy_MC}
    E_{MC} = A \sum_i (E_{scint,i} + E_{c,i}),
\end{equation}

where \(A\) is a scaling factor adjusting the overall energy-to-light conversion efficiency.

\subsection{\pspt~Gamma Simulation}
\label{sec:prospect_gamma_sim}
To validate the unfolded gamma source term models from Section~\ref{sec:detector_response} and explore the use of reactor neutron gamma capture peaks for calibration purposes in future experiments, a simulation of the gamma spectrum as taken at the Shield Center location (near the northeast corner of the PROSPECT boundary), pictured in Figure~\ref{fig:prospect_boundary_gamma}, was created.
This was done by unfolding the spectrum using the detector response as detailed in Section~\ref{sec:detector_response} and feeding this gamma flux spectrum into the \pspt~simulation package.

The Shield Center spectrum was chosen because it was taken close to the reactor and therefore contains some of the \el{Ni}{59}~gammas. It was observed within the \pspt~data that a broad peak around 11.39 MeV was visible, but only in a few segments near the North face of the detector and close to the floor (recall that the reactor is situated $\approx$5~m below the experiment hall). Thus, the Shield Center data was most representative of the gammas seen in those segments.

The simulation generates an isotropic flux of gammas utilizing an unfolded spectrum from the Ge dataset for the energy distribution. The full AD1 geometry, encompassing the detector and shield package, was utilized. For computational efficiency, momenta with a uniform distribution were sampled and trajectories that did not intersect with the scintillator volume were not propagated in the simulation.

\begin{figure*}[pos=htbp]
    \centering
    \includegraphics[width=1.0\linewidth]{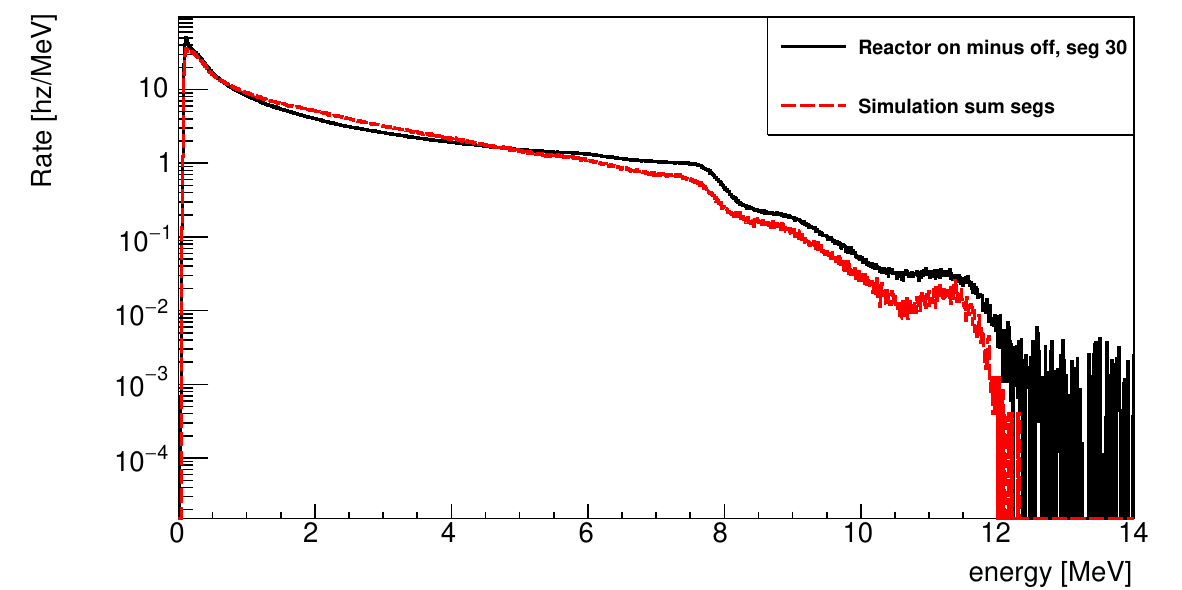}
    \caption{Data (black) vs simulation (red) for gamma backgrounds in the \pspt~detector. The data is gamma energy spectrum in segment 30 from 3 weeks of reactor on data taken in September 2018 with a background subtraction performed using 1 week of reactor off data to remove cosmic backgrounds. Simulation uses the unfolded spectrum measured with the collimated Ge detector at the Shield Center location as input. Simulation is normalized to minimize the chi-square difference with the data over the full plotted energy range in order to compare spectral shape. Note the simulation uses the sum of all detector segments spectra in order to save computational time; energy leakage effects are expected to be more significant for a single segment than for the sum of all segments.}
    \label{fig:prospect_nw_sim}
\end{figure*}

The result of this simulation is shown in Figure~\ref{fig:prospect_nw_sim}. This simulation uses the unfolded `SHIELD CENTER' spectrum shown in Figure~\ref{fig:HB4_shield}, taken at the brown-star Shield Center location pictured in Figure~\ref{fig:key_measurement_locations}. The simulation result is normalized to minimize the chi-square difference with the data over the full plotted energy range (0.1--13 MeV) in order to give a comparison between spectral shapes.

Features visible in the data energy spectrum are the Compton edges at $\approx$ 7.8 MeV from neutron capture on Fe and Al, a broader Compton edge at $\approx$ 9 MeV from the higher energy Fe and Ni neutron captures, and the Gaussian feature at 11.4 MeV from neutron capture on \el{Ni}{59}. The next Section details the results of this simulation-data comparison.

\subsection{Simulation Comparison}
\label{sec:ni_59_fit_results}

A Gaussian plus a linear polynomial was fit to both the data and simulation for segments 15 and 30 for the energy range 10.0 to 12.0 MeV. These segments were chosen because they are the closest segments to the reactor that had working PMTs on both ends of the segments for the duration of the neutrino measurement campaign (see Figure~\ref{fig:prospect_IBD_efficiency} for the numbering scheme of the \pspt~segments, note that the reactor is down and to the left of this Figure).

A comparison between the measured data and Monte Carlo (MC) simulations for representative front-bottom segments 15, 16, and 30 shows reasonable overall agreement across the energy spectrum. Specifically, from approximately 200 keV to 1 MeV, the data and MC concur within better than 10\%. Between 1 MeV and 6 MeV, the agreement remains within about 25\%, while from 6 MeV to 11.5 MeV, discrepancies are limited to a factor of approximately 2.

We were unable to fit a bump at 11.4 MeV for Segment 16, due to insufficient full energy deposits of the 11.4 MeV gamma within that segment to form an appreciable Gaussian-like feature. However, both segments 15 and 30 spectra resulted in fits with centroids matching the simulated energy spectrum.

Studies performed with multi-week PROSPECT datasets indicate that week-scale exposures may already provide a useful calibration cross-check at these energies, but the present study does not quantify a precise one-week sensitivity threshold. In addition, the resolution for the highest-energy feature does not agree between data and simulation, and a simple common Gaussian resolution smearing cannot reconcile differences for all investigated PROSPECT front-bottom segments. Several factors may contribute to this discrepancy: (a) the parametrized light yield and quenching model in PG4 is only well-validated in the energy range below approximately 8 MeV where PROSPECT antineutrino measurements were made; (b) the Cherenkov light contribution at high energies may not be accurately captured by the phenomenological model described in Section~\ref{sec:PG4}; (c) position-dependent light collection effects within the elongated scintillator segments; and (d) energy leakage between adjacent segments, which is more significant for the 11.4 MeV gamma than for lower-energy calibration sources. Addressing these factors would help guide the calibration strategy for future experiments.

\section{Conclusion}

This work characterized the gamma radiation field in the HFIR Experiment Hall and identified the primary sources and their spatial distribution. The HB4 source was shown to be the largest component contributing to radiation at the location of the \pspt~detector after the local shield wall was placed, which mitigated gammas coming directly from the reactor as demonstrated through quantitative comparison of measurements at the MIF location and behind the lead shield wall. Gamma backgrounds remain low above the monolith in the 1-2 feet to the east of the PROSPECT location, but increase further east and south beyond this due to backgrounds from the HB4 beamline.  Gamma backgrounds remain low 1-2 feet to the west and many feet to the southwest of PROSPECT, with background increasing beyond the end of the local shield wall.

HPGe measurements were unfolded to provide gamma flux models at various locations that may be useful in planning and modeling backgrounds for future particle physics efforts, such as reactor \cevns~or beyond-standard-model particle search experiments. Both response-model limits (isotropic and front-face), together with calibrated measured spectra and metadata tables, are provided in tabulated format in the supplemental materials.

Representative validation of gamma flux models with \pspt~datasets showed agreement with spectral shape to better than 10\% from roughly 200 keV to 1 MeV, within about 25\% from 1-6 MeV, and within a factor of 2 from 6-11.5 MeV for the segments studied here. This demonstrates sufficient fidelity of the gamma source models for future experimental planning, though resolution modeling at high energies requires further development.
It is important to note that while spectral shape agreement is good, the absolute flux normalization from the unfolded spectrum carries additional systematic uncertainty from the detector-response simulation used to construct the migration matrix. This includes the assumed illumination geometry, but also the modeled transport through the detector, collimator, shielding, and surrounding materials. These effects are difficult to model accurately given the inability of the collimator to shield high-energy gammas effectively (see Appendix~\ref{app:validation}). We therefore treat the isotropic response and the uniform front-face response as limiting cases that bracket part of the plausible flux range, while recognizing that the remaining low-energy suppression in the front-face case points to additional detector-plus-shield response mismodeling. An alternate penalized-Poisson unfolding crosscheck and toy counting-statistics study support this interpretation: for the reactor-facing MIF case, the low-energy suppression persists across unfolding algorithms, so the dominant issue is not the Richardson-Lucy method itself. For some locations, such as in the vicinity of the Russian Doll setup behind the shield wall, the anisotropic flux model yielded robust unfolding results at all energies, indicating a predominantly isotropic flux at these locations.

Investigation into background reduction using Russian doll shielding showed that thermal neutrons become the primary background source at low energies once gammas are mitigated with sufficient shielding. With approximately 23~cm of water shielding surrounding the shield, reactor-related backgrounds approach within a factor of 2-4 of ambient backgrounds in the 30-60 keV energy range, with measured reactor-on rates of approximately 28.3 mHz/kg/keV (see Table~\ref{tab:rd_rates}). This indicates that the Russian doll configuration approaches the regime in which cosmic rays and material radioactivity dominate, making it potentially suitable for \cevns~detection applications. 

A detailed characterization of the thermal neutron spectrum and its variations with experimental activity would be required to determine the feasibility of future \cevns~measurements at the HFIR Experiment Hall. Gamma backgrounds at higher energies (0.5-2.0 MeV) are about 4.5 times those of reactor off rates, indicating that further mitigation strategies are needed for other beyond-standard-model particle searches relevant in this energy range. 

\clearpage
\appendix
\section{Gamma Spectral Lines at HFIR Experiment Hall}

\label{sec:spectral_lines}
\begin{figure*}[pos=!bp]
    \centering
    \includegraphics[width=0.7\linewidth]{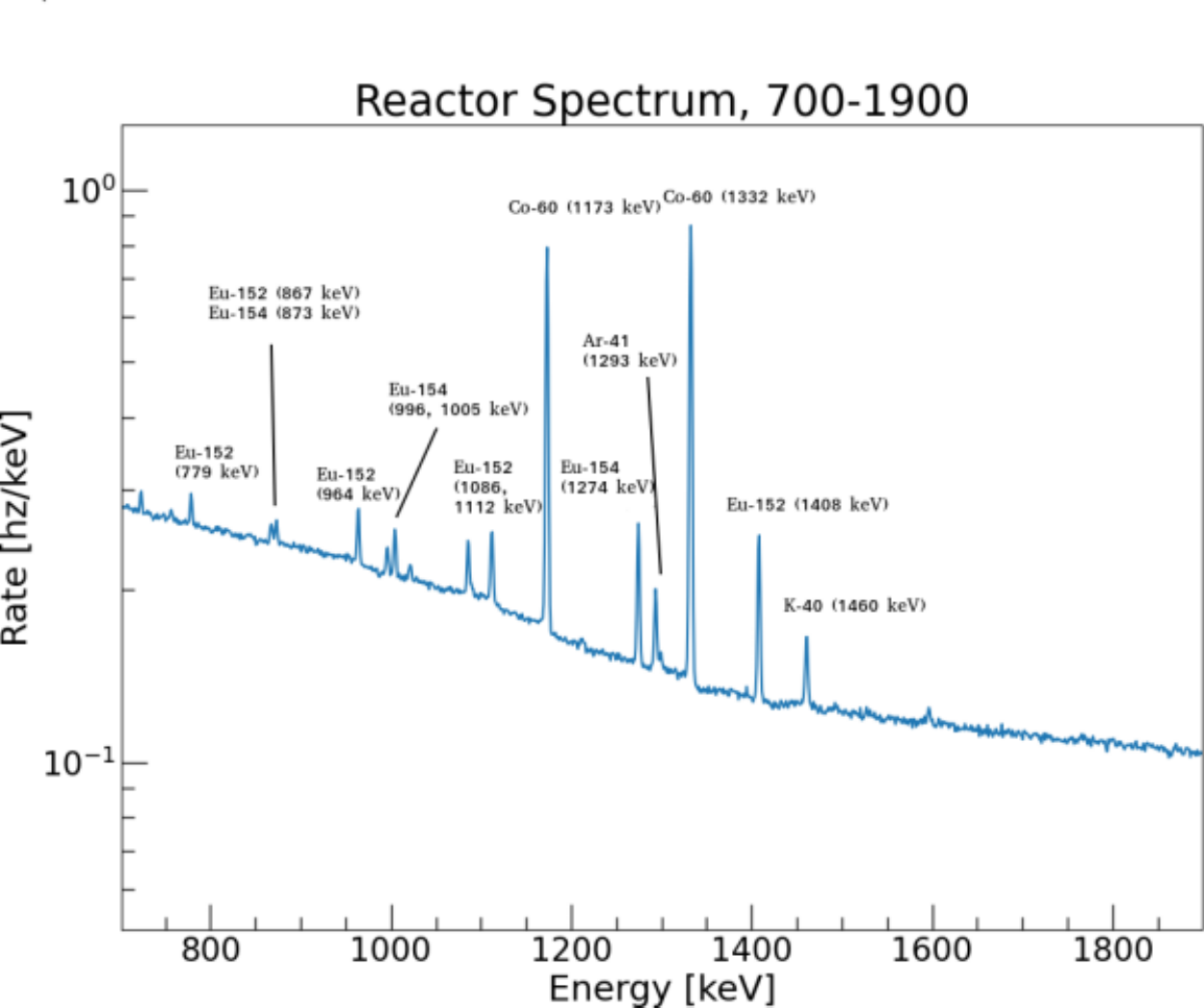}
    \caption{Peaks labelled for spectrum taken under the MIF with collimator pointed at the reactor core, 700 - 1900 keV.}
    \label{fig:labelled_spectra_1}
\end{figure*}

\begin{figure*}[pos=!bp]
    \centering
    \includegraphics[width=0.7\linewidth]{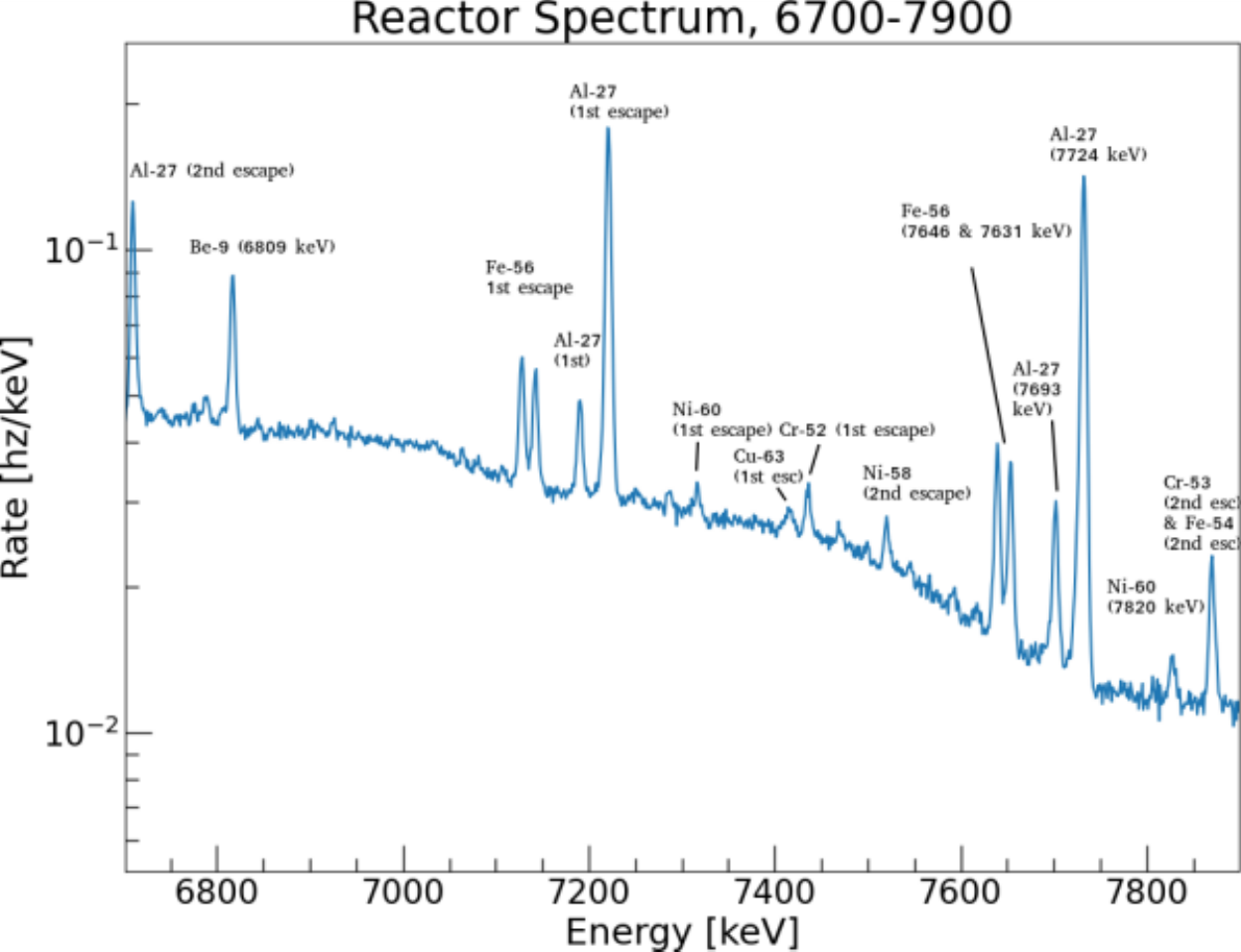}
    \caption{Peaks labelled for spectrum taken under the MIF with collimator pointed at the reactor core, 6700 - 7900 keV.}
    \label{fig:labelled_spectra_2}
\end{figure*}

\begin{figure*}[pos=!bp]
    \centering
    \includegraphics[width=0.7\linewidth]{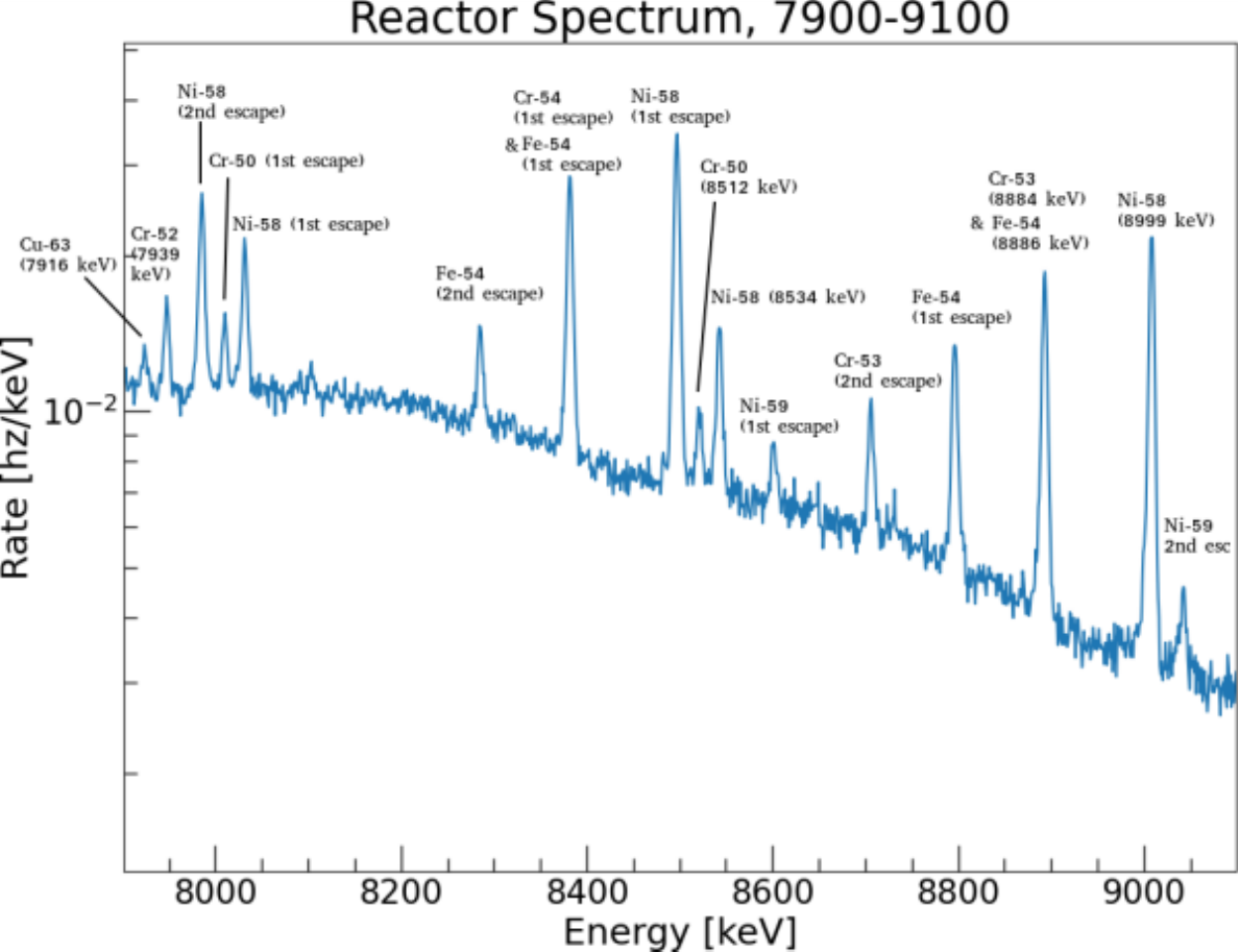}
    \caption{Peaks labelled for spectrum taken under the MIF with collimator pointed at the reactor core, 7900 - 9100 keV.}
    \label{fig:labelled_spectra_3}
\end{figure*}

\begin{figure*}[pos=!bp]
    \centering
    \includegraphics[width=0.7\linewidth]{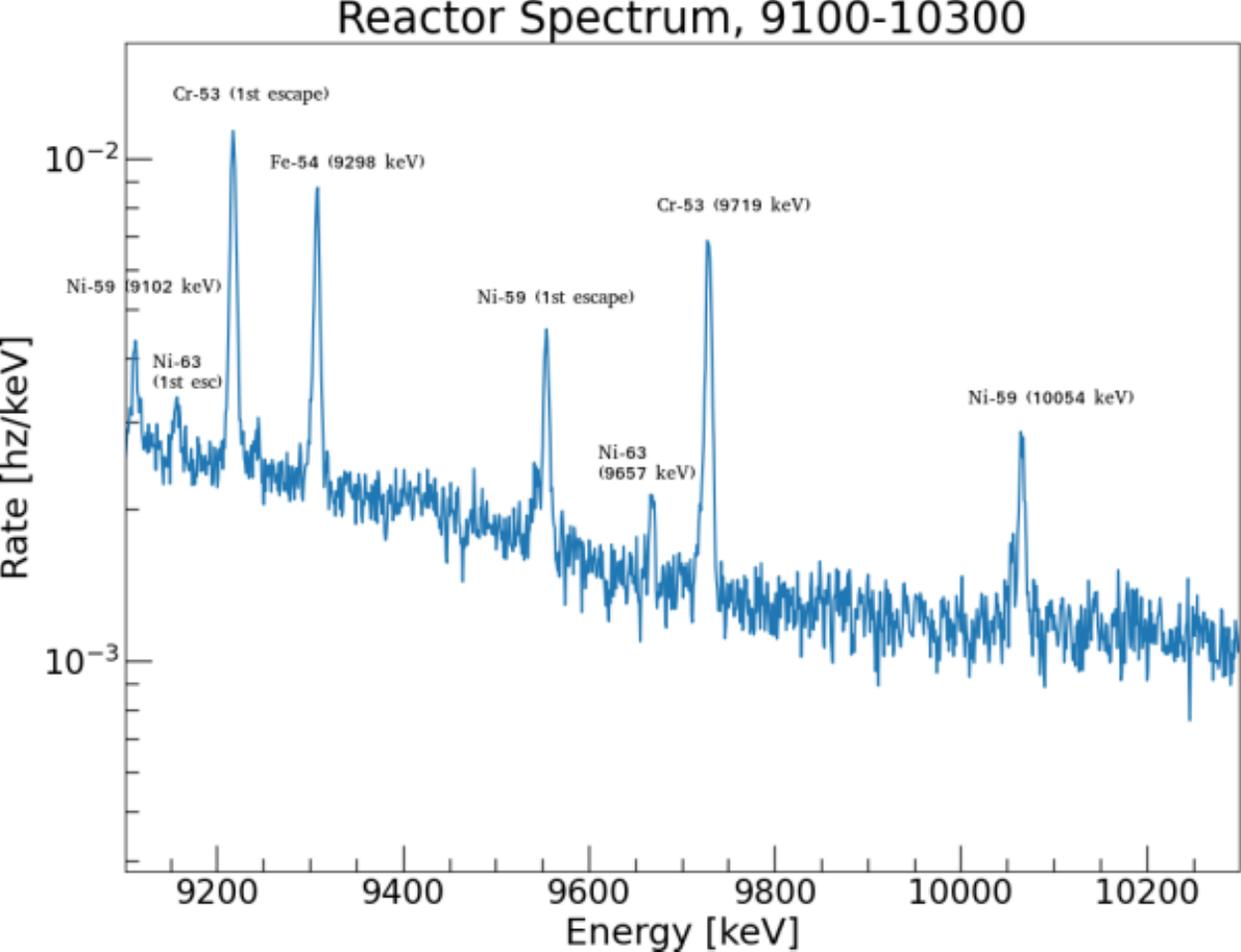}
    \caption{Peaks labelled for spectrum taken under the MIF with collimator pointed at the reactor core, 9100 - 10300 keV.}
    \label{fig:labelled_spectra_4}
\end{figure*}

\begin{figure*}[pos=!bp]
    \centering
    \includegraphics[width=0.7\linewidth]{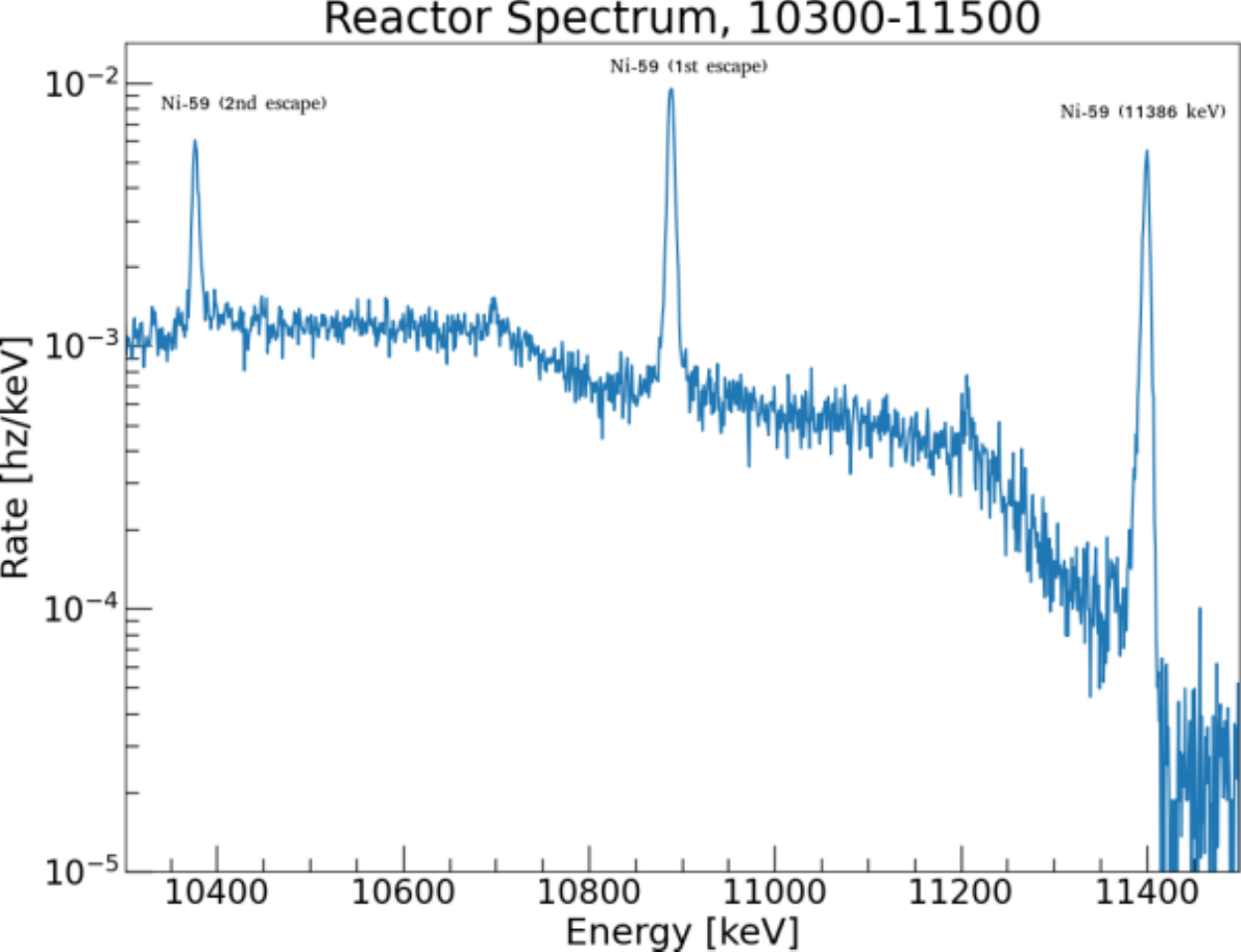}
    \caption{Peaks labelled for spectrum taken under the MIF with collimator pointed at the reactor core, 10300 - 11500 keV.}
    \label{fig:labelled_spectra_5}
\end{figure*}

A list of gammas and their corresponding parent isotopes found in the gamma spectra
taken in the experiment hall at HFIR
is shown in table~\ref{table:source_catalogue}. Gamma flux from all isotopes in Table~\ref{table:source_catalogue} are present and come from the direction of the reactor with the exception of $^{48}\textrm{Ti}$ gammas which are present in spectra coming from the floor above the HB4 hot spot. Energies listed are from the most frequent gammas resulting from neutron capture, except for Eu, Ar, Co, and K, which are decays from radioactive isotopes. Energies and yields are obtained from the nudat database~\cite{nudat} for radioactive decays and capgam~\cite{capgam} for neutron capture reactions.
What follows is the gamma spectrum depicted in Figure~\ref{fig:reactor_spectrum} broken up into various energy ranges with the peaks identified in Figures~\ref{fig:labelled_spectra_1}--\ref{fig:labelled_spectra_5}.

\clearpage
\begin{table*}[pos=!bp]
    \caption{Gamma lines and yields~\cite{nndc} associated with spectral peaks acquired with the Ge detector with the collimator pointing in the direction of the reactor or the HB4 hotspot. Columns 4 and 5 provide the reaction and origin or source of each line.}\label{table:source_catalogue}
    \centering
    \footnotesize
    \setlength{\tabcolsep}{4pt}
    \begin{tabular}{llllll}
        \toprule
         & isotope             & energy [keV] & yield [\%]~\cite{nndc} & reaction          & source \\
        \midrule
         & $^{154}\textrm{eu}$ & 723.30       & 20.06      & radioactive decay & Eu in reactor control plates/pool \\
         & $^{152}\textrm{eu}$ & 778.9        & 12.93      & radioactive decay & Eu in reactor control plates/pool \\
         & $^{152}\textrm{eu}$ & 841.63       & 14.2       & radioactive decay & Eu in reactor control plates/pool \\
         & $^{9}\textrm{be}$   & 853.63       & 35.812     & neutron capture   & Be reflector near reactor core \\
         & $^{152}\textrm{eu}$ & 963.38       & 11.6       & radioactive decay & Eu in reactor control plates/pool \\
         & $^{152}\textrm{eu}$ & 964.01       & 14.51      & radioactive decay & Eu in reactor control plates/pool \\
         & $^{152}\textrm{eu}$ & 1085.84      & 10.11      & radioactive decay & Eu in reactor control plates/pool \\
         & $^{152}\textrm{eu}$ & 1112.08      & 13.67      & radioactive decay & Eu in reactor control plates/pool \\
         & $^{60}\textrm{co}$  & 1173.2       & 99.85      & radioactive decay & residual Co near MIF valve box \\
         & $^{154}\textrm{eu}$ & 1274.43      & 34.83      & radioactive decay & Eu in reactor control plates/pool \\
         & $^{41}\textrm{ar}$  & 1293.64      & 99.16      & radioactive decay & activated air circulating during reactor-on \\
         & $^{60}\textrm{co}$  & 1332.5       & 99.98      & radioactive decay & residual Co near MIF valve box \\
         & $^{24}\textrm{na}$ & 1368.63 & 99.99 & radioactive decay & activated Na in building materials \\
         & $^{152}\textrm{eu}$ & 1408.01      & 20.87      & radioactive decay & Eu in reactor control plates/pool \\
         & $^{40}\textrm{k}$   & 1460.82      & 10.66      & radioactive decay & natural K in building materials \\
         & $^{9}\textrm{be}$   & 2590.014     & 32.929     & neutron capture   & Be reflector near reactor core \\
         & $^{208}\textrm{tl}$ & 2614.51 & 99.75 & radioactive decay & Th-series in building materials \\
         & $^{24}\textrm{na}$ & 2754.01 & 99.87 & radioactive decay & activated Na in building materials \\
         & $^{9}\textrm{be}$   & 3367.45      & 49.165     & neutron capture   & Be reflector near reactor core \\
         & $^{9}\textrm{be}$   & 3443.406     & 16.84      & neutron capture   & Be reflector near reactor core \\
         & $^{52}\textrm{cr}$  & 5269         & 11.331     & neutron capture   & structural steel near reactor/HB4 \\
         & $^{52}\textrm{cr}$  & 5618.23      & 32.287     & neutron capture   & structural steel near reactor/HB4 \\
         & $^{56}\textrm{fe}$  & 5920.35      & 33.103     & neutron capture   & structural steel near reactor/HB4 \\
         & $^{56}\textrm{fe}$  & 6018.42      & 34.138     & neutron capture   & structural steel near reactor/HB4 \\
         & $^{50}\textrm{cr}$  & 6135.9       & 11.263     & neutron capture   & structural steel near reactor/HB4 \\
         & $^{48}\textrm{ti}$  & 6418.53      & 35.67      & neutron capture   & HB4 beamline shielding \\
         & $^{53}\textrm{cr}$  & 6645.64      & 12.291     & neutron capture   & structural steel near reactor/HB4 \\
         & $^{48}\textrm{ti}$  & 6760.12      & 54.15      & neutron capture   & HB4 beamline shielding \\
         & $^{9}\textrm{be}$   & 6809.61      & 100        & neutron capture   & Be reflector near reactor core \\
         & $^{53}\textrm{cr}$  & 7100.11      & 9.633      & neutron capture   & structural steel near reactor/HB4 \\
         & $^{63}\textrm{cu}$  & 7253.05      & 12.538     & neutron capture   & Cu in facility materials \\
         & $^{56}\textrm{fe}$  & 7278.82      & 20.69      & neutron capture   & structural steel near reactor/HB4 \\
         & $^{63}\textrm{cu}$  & 7307.31      & 27.0695    & neutron capture   & Cu in facility materials \\
         & $^{50}\textrm{cr}$  & 7362.6       & 15.525     & neutron capture   & structural steel near reactor/HB4 \\
         & $^{52}\textrm{cr}$  & 7374.58      & 18.942     & neutron capture   & structural steel near reactor/HB4 \\
         & $^{60}\textrm{ni}$  & 7536.62      & 57.0388    & neutron capture   & structural steel near reactor/HB4 \\
         & $^{56}\textrm{fe}$  & 7631.18      & 100        & neutron capture   & structural steel near reactor/HB4 \\
         & $^{63}\textrm{cu}$  & 7638         & 48.943     & neutron capture   & Cu in facility materials \\
         & $^{56}\textrm{fe}$  & 7645.58      & 86.207     & neutron capture   & structural steel near reactor/HB4 \\
         & $^{27}\textrm{al}$  & 7693.398     & 12.366     & neutron capture   & reactor vessel Al \\
         & $^{27}\textrm{al}$  & 7724.034     & 96.0573    & neutron capture   & reactor vessel Al \\
         & $^{60}\textrm{ni}$  & 7819.56      & 100        & neutron capture   & structural steel near reactor/HB4 \\
         & $^{63}\textrm{cu}$  & 7916.26      & 100        & neutron capture   & Cu in facility materials \\
         & $^{52}\textrm{cr}$  & 7938.58      & 100        & neutron capture   & structural steel near reactor/HB4 \\
         & $^{58}\textrm{ni}$  & 8120.75      & 8.501      & neutron capture   & structural steel near reactor/HB4 \\
         & $^{63}\textrm{ni}$  & 8311.45      & 9.825      & neutron capture   & structural steel near reactor \\
         & $^{50}\textrm{cr}$  & 8484.2       & 32.5       & neutron capture   & structural steel near reactor/HB4 \\
         & $^{50}\textrm{cr}$  & 8512.2       & 37.5       & neutron capture   & structural steel near reactor/HB4 \\
         & $^{58}\textrm{ni}$  & 8533.71      & 47.839     & neutron capture   & structural steel near reactor/HB4 \\
         & $^{53}\textrm{cr}$  & 8884.81      & 56.203     & neutron capture   & structural steel near reactor/HB4 \\
         & $^{54}\textrm{fe}$  & 8886.4       & 18.636     & neutron capture   & structural steel near reactor/HB4 \\
         & $^{58}\textrm{ni}$  & 8998.63      & 100        & neutron capture   & structural steel near reactor/HB4 \\
         & $^{59}\textrm{ni}$  & 9102.1       & 19.798     & neutron capture   & structural steel near reactor core \\
         & $^{54}\textrm{fe}$  & 9297.8       & 100        & neutron capture   & structural steel near reactor/HB4 \\
         & $^{63}\textrm{ni}$  & 9656.89      & 80.702     & neutron capture   & structural steel near reactor \\
         & $^{53}\textrm{cr}$  & 9718.79      & 20         & neutron capture   & structural steel near reactor/HB4 \\
         & $^{59}\textrm{ni}$  & 10054.14     & 18.43      & neutron capture   & structural steel near reactor core \\
         & $^{59}\textrm{ni}$  & 11386.5      & 48.206     & neutron capture   & structural steel near reactor core \\
        \bottomrule
    \end{tabular}
\end{table*}


\clearpage
\section{Supplemental Validation Data}
\label{app:validation}
\subsection{Simulation Validation at High Energy}
Without known rates of high energy sources we cannot measure the detector efficiency at these energies, but we can measure the ratio of the full peak area to the first and second escape peaks.
These ratios should be captured accurately by the simulation if the geometry is correct and the Compton scattering, photoelectric effect,
and pair production physics are accurately reproduced. The area of peaks were measured by taking the background from the fit used in equation~\ref{eq:peak_fit} and subtracting this from the histogram values 3.5 $\sigma$ to the left and the right of the peak centroid.
Linear interpolation between samples was used to approximate partial bin area. The spectrum used for this study was the same reactor on spectrum as in Figure~\ref{fig:reactor_spectrum}. The results are shown in table~\ref{table:peak_ratio_compare}.
In general, the ratios agree better for the isotropic response-model limit than for the front-face limit. The next Section compares these two limiting detector-response models directly and shows that the front-face case produces a much larger source-area-normalized photopeak response at low energy, while the separation becomes more modest in the multi-MeV region.

\clearpage
\begin{table*}[pos=!bp]
    \caption{Peak ratios for selected neutron capture gammas taken at the MIF reactor-facing position. 0, 1, and 2 are the areas of the main, first escape, and second escape peaks respectively. In general the ratios are better reproduced by the isotropic response-model limit than by the front-face limit. E = energy, R = ratio, dR = ratio uncertainty.}
    \label{table:peak_ratio_compare}
    \begin{tabular}{lllllllll}
        \toprule
         & E {[}keV{]} & type        & R 0/1 & dR 0/1 & R 0/2 & dR 0/2 & R 1/2 & dR 1/2 \\
        \midrule
& 6809.61     & data        & 1.02  & 1.93   & 2.13  & 6.83   & 2.08  & 6.88   \\
& 6809.61     & sim through & 1.34  & 0.09   & 2.42  & 0.20   & 1.81  & 0.19   \\
& 6809.61     & sim all     & 1.05  & 0.07   & 1.70  & 0.17   & 1.63  & 0.20   \\
& 7631.18     & data        & 0.90  & 1.44   & 1.83  & 7.09   & 2.04  & 8.30   \\
& 7631.18     & sim through & 1.14  & 0.07   & 2.13  & 0.21   & 1.87  & 0.22   \\
& 7631.18     & sim all     & 0.87  & 0.05   & 1.49  & 0.16   & 1.70  & 0.21   \\
& 7645.58     & data        & 0.86  & 1.56   & 1.53  & 6.45   & 1.78  & 7.90   \\
& 7645.58     & sim through & 1.16  & 0.09   & 2.17  & 0.24   & 1.88  & 0.25   \\
& 7645.58     & sim all     & 0.89  & 0.05   & 1.48  & 0.15   & 1.67  & 0.20   \\
& 7693.40     & data        & 0.99  & 2.40   & 1.52  & 7.28   & 1.55  & 7.98   \\
& 7693.40     & sim through & 1.15  & 0.08   & 2.08  & 0.20   & 1.81  & 0.21   \\
& 7693.40     & sim all     & 0.86  & 0.06   & 1.45  & 0.14   & 1.68  & 0.19   \\
& 7724.03     & data        & 0.87  & 0.27   & 1.85  & 1.49   & 2.12  & 1.79   \\
& 7724.03     & sim through & 1.11  & 0.07   & 2.09  & 0.21   & 1.88  & 0.23   \\
& 7724.03     & sim all     & 0.85  & 0.05   & 1.49  & 0.15   & 1.75  & 0.20   \\
& 8998.63     & data        & 0.67  & 0.28   & 1.36  & 1.41   & 2.04  & 2.20   \\
& 8998.63     & sim through & 0.97  & 0.08   & 1.88  & 0.26   & 1.93  & 0.30   \\
& 8998.63     & sim all     & 0.75  & 0.05   & 1.24  & 0.14   & 1.65  & 0.22   \\
& 9718.79     & data        & 0.72  & 0.35   & 1.25  & 1.96   & 1.74  & 2.77   \\
& 9718.79     & sim through & 0.86  & 0.05   & 1.68  & 0.22   & 1.95  & 0.28   \\
& 9718.79     & sim all     & 0.67  & 0.04   & 1.16  & 0.16   & 1.73  & 0.26   \\
& 11386.50    & data        & 0.58  & 0.07   & 1.28  & 0.50   & 2.19  & 0.90   \\
& 11386.50    & sim through & 0.77  & 0.05   & 1.49  & 0.26   & 1.95  & 0.37   \\
& 11386.50    & sim all     & 0.59  & 0.04   & 1.08  & 0.19   & 1.84  & 0.34   \\
        \bottomrule
    \end{tabular}
\end{table*}

\subsection{Collimator Effectiveness}
\label{collimator_effectiveness}
\begin{figure*}[pos=htbp]
\centering
\includegraphics[width=0.75\textwidth]{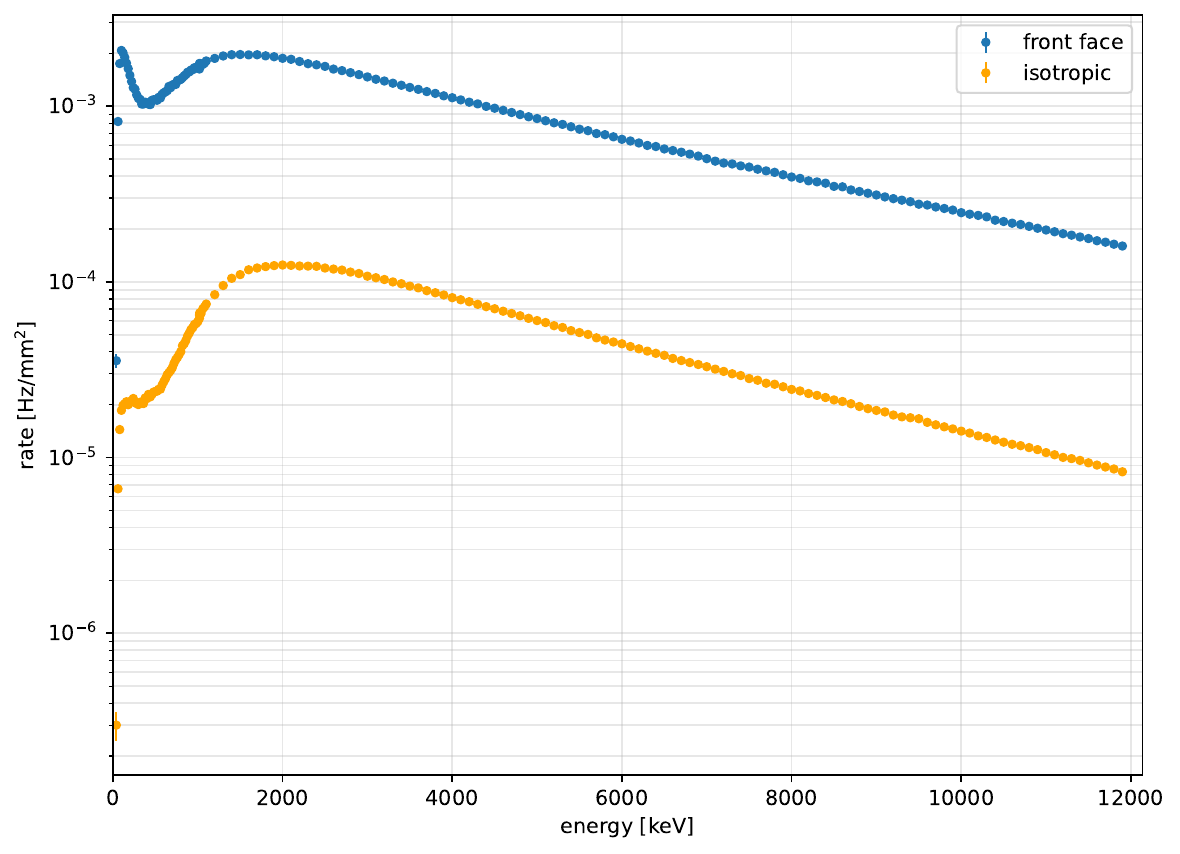}
\caption{Detected photopeak rates from corrected simulations for the two response-model limits used in the unfolding: an isotropic 1~Hz/mm$^2$ gamma flux over the outer surface of the collimator-detector system, and a uniform 1~Hz/mm$^2$ gamma flux directed into the detector-collimator front face. The rates are divided by the corresponding source surface area so that the two curves can be compared as relative detection probabilities per incident gamma.}
\label{fig:collimator_effectiveness}
\end{figure*}

To compare the two limiting detector-response models used in this work, simulations were performed at various energies for two source geometries: an isotropic flux over the outer surface of the collimator-detector system, and a uniform flux directed into the detector-collimator front face. Figure~\ref{fig:collimator_effectiveness} plots the detected photopeak rate for each case after normalizing by the corresponding source surface area, so the ratio of the curves can be interpreted as the relative detection probability per incident gamma under the two angular assumptions. The front-face case gives a much larger normalized response at low energies, reaching about a factor of 120 above the isotropic case at 60--80~keV, while the separation falls to about 15 at 2000~keV and stays in the 14--19 range through the rest of the multi-MeV region. This behavior is consistent with the lead collimator strongly suppressing off-axis low-energy gammas while becoming less selective once higher-energy gammas can penetrate the lead from shielded directions.

\clearpage
\section*{Acknowledgments}
This material is based upon work supported by the following sources: US Department of Energy (DOE) Office of Science, Office of High Energy Physics under Award No. DE-SC0016357 and DE-SC0017660 to Yale University, under Award No. DE-SC0017815 to Drexel University, under Award No. DE-SC0008347 to Illinois Institute of Technology, under Award No. DE-SC0010504 to University of Hawaii, under Contract No. DE-SC0012704 to Brookhaven National Laboratory, and under Work Proposal Number  SCW1504 to Lawrence Livermore National Laboratory.
This work was performed under the auspices of the U.S. Department of Energy by Lawrence Livermore National Laboratory under Contract DE-AC52-07NA27344 and by Oak Ridge National Laboratory under Contract DE-AC05-00OR22725.
Additional funding for the experiment was provided by the Heising-Simons Foundation under Award No. \#2016-117 to Yale University.

We further acknowledge support from Yale University, the Illinois Institute of Technology, Temple University, University of Hawaii, Brookhaven National Laboratory, the Lawrence Livermore National Laboratory LDRD program, the National Institute of Standards and Technology, and Oak Ridge National Laboratory.
Illinois Institute of Technology efforts were also sponsored by the Defense Advanced Research Projects Agency under cooperative agreement HR0011-25-2-0035.
We gratefully acknowledge the support and hospitality of the High Flux Isotope Reactor and Oak Ridge National Laboratory, managed by UT-Battelle for the U.S. Department of Energy.

The views expressed in this paper are those of the authors and do not reflect the official policy or position of the U.S. Naval Academy, the Department of Defense, or the U.S. Government, and no official endorsement should be inferred.